\documentclass[aps,twocolumn,nofootinbib, superscriptaddress, floatfix]{revtex4-1}

\usepackage{placeins}
\usepackage{soul}
\usepackage{amsmath}
\usepackage[usenames,dvipsnames,svgnames]{xcolor}
\definecolor{redd}{rgb}{0.8, 0.1,0.2}
\definecolor{navy}{rgb}{0.05, 0.23,0.75}
\usepackage[colorlinks]{hyperref}
\usepackage{chngcntr}
\hypersetup{
     colorlinks   = true,
     citecolor    = navy,
	linkcolor = redd,
	urlcolor=navy,
	anchorcolor=blue
}
\usepackage{bbm}
\usepackage{amsfonts}
\usepackage{amsmath}
\usepackage{mathrsfs}
\usepackage{soul}

\usepackage{amsmath,amssymb,amsbsy,bm}

\usepackage{scalerel,tikz}
\usetikzlibrary{svg.path}
\definecolor{orcidlogocol}{HTML}{A6CE39}
\tikzset{orcidlogo/.pic={
 \fill[orcidlogocol] svg{M256,128c0,70.7-57.3,128-128,128C57.3,256,0,198.7,0,128C0,57.3,57.3,0,128,0C198.7,0,256,57.3,256,128z};
 \fill[white] svg{M86.3,186.2H70.9V79.1h15.4v48.4V186.2z}
 svg{M108.9,79.1h41.6c39.6,0,57,28.3,57,53.6c0,27.5-21.5,53.6-56.8,53.6h-41.8V79.1z M124.3,172.4h24.5c34.9,0,42.9-26.5,42.9-39.7c0-21.5-13.7-39.7-43.7-39.7h-23.7V172.4z}
 svg{M88.7,56.8c0,5.5-4.5,10.1-10.1,10.1c-5.6,0-10.1-4.6-10.1-10.1c0-5.6,4.5-10.1,10.1-10.1C84.2,46.7,88.7,51.3,88.7,56.8z};
}}
\newcommand\orcidicon[1]{\href{https://orcid.org/#1}{\mbox{\scalerel*{
\begin{tikzpicture}[yscale=-1,transform shape]
\pic{orcidlogo};
\end{tikzpicture}
}{|}}}}

\usepackage{epsfig}
\usepackage{graphicx}
\usepackage{wrapfig}                 
\usepackage{url}
\usepackage{hyperref}
\usepackage{float}
\usepackage{pstricks}
\usepackage{color}
\usepackage{multirow}
\usepackage{lipsum}
\usepackage{enumitem}
\usepackage{ctable}
\newcolumntype{L}{>{\centering\arraybackslash}m{1.5cm}}

\usepackage{enumitem}

\usepackage{chngcntr}

\newcommand{\be}{\begin{equation}}
\newcommand{\ee}{\end{equation}}
\newcommand{\bea}{\begin{eqnarray}}
\newcommand{\Beq}{\begin{equation}\begin{aligned}}
\newcommand{\Eeq}{\end{aligned}\end{equation}}
\newcommand{\eea}{\end{eqnarray}}
\newcommand{\beq}{\begin{equation}}
\newcommand{\eeq}{\end{equation}}
\newcommand{\bk}{{\bf k}}
\newcommand{\bq}{{\bf q}}
\newcommand{\bx}{{\bf x}}

\newcommand{\mpl}{{m_{\rm pl}}}

\newcommand{\aap}{AAP}
\newcommand{\mnras}{Mon. Not. Roy. Astr. Soc.}

\definecolor{cerulean}{rgb}{0., 0.62,0.7}

\newcommand{\twiddle}{{\raise.17ex\hbox{$\scriptstyle\sim$}}}

\begin{document}

\title{Formation, Gravitational Clustering and Interactions of \\non-relativistic Solitons in an Expanding Universe}
\author{Mustafa A. Amin\,\orcidicon{0000-0002-8742-197X}}
\email{mustafa.a.amin@gmail.com} 
\affiliation{Physics \& Astronomy Department, Rice University, Houston, Texas 77005, USA}
\author{Philip Mocz\,\orcidicon{0000-0001-6631-2566}}
\email{philip.mocz@gmail.com} 
\thanks{Einstein Fellow}
\affiliation{Department of Astrophysical Sciences, Princeton University, 4 Ivy Lane, Princeton, NJ, 08544, USA}

\begin{abstract}
We investigate the formation, gravitational clustering, and interactions of solitons in a self-interacting, non-relativistic scalar field in an expanding universe. Rapid formation of a large number of solitons is driven by attractive self-interactions of the field, whereas  the slower clustering of solitons is driven by gravitational forces. Driven closer together by gravity, we see a rich plethora of dynamics in the soliton ``gas'' including mergers, scatterings and formation of soliton binaries. The numerical simulations are complemented by analytic calculations and estimates of (i) the relevant instability length and time scales, (ii) individual soliton profiles and their stability,  (iii) number density of produced solitons, and (iv) the two-point correlation function of soliton positions as evidence for gravitational clustering. 
\end{abstract}

\maketitle

\section{Introduction}
Solitons are self-localized, persistent configurations in nonlinear field theories which have been studied intensely in a broad range of contexts including cosmology, high energy physics, nonlinear optics and cold-atom physics, condensed matter physics, fluid mechanics and mathematics \cite{Lee:1991ax,Vilenkin:2000jqa,Manton:2004tk,Weinberg:2012pjx,Liebling:2012fv,RevModPhys.61.763}. 

In cosmology,  for example, solitons can emerge naturally at the end of inflation and dominate the energy density (e.g.~\cite{Amin:2011hj}), or related configurations can form in the axion field that might constitute the entirety or part of the dark matter (e.g.~\cite{Kolb:1993hw}). Depending on the context, they can act as new sources of gravitational waves \cite{Zhou:2013tsa,Antusch:2017flz,Helfer:2018vtq,Palenzuela:2017kcg,Lozanov:2019ylm}, potentially lead to the formation of primordial  black holes \cite{Khlopov:1985jw,Cotner:2016cvr,Cotner:2018vug}, be involved in baryogenesis \cite{Enqvist:1997si,LozAmin}, and change the approach to radiation domination in the early universe \cite{Adshead:2015pva,Lozanov:2016hid,Lozanov:2017hjm}; they can also provide novel insights into the small scale problems in the cold dark matter paradigm \cite{Hu:2000ke,Alcubierre:2001ea,Marsh:2015wka,Schive:2014dra,Hui:2016ltb}. 

\begin{figure*}[t!]
\begin{tabular}{ccccccc}
$a=$ & $1$ & $2$ & $4$ & $8$ & $16$ & $20$ \\
\rotatebox{90}{\,\,\,\,\,\,\,\,\,\,w/ gravity} & 
\includegraphics[width=0.15\textwidth]{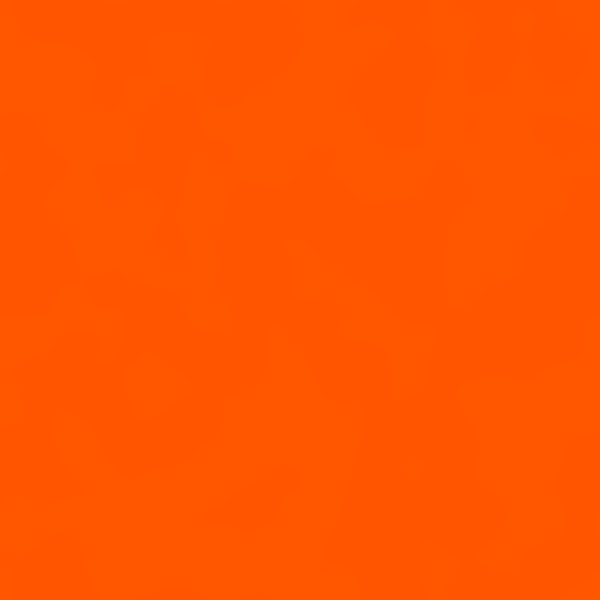}  &
\includegraphics[width=0.15\textwidth]{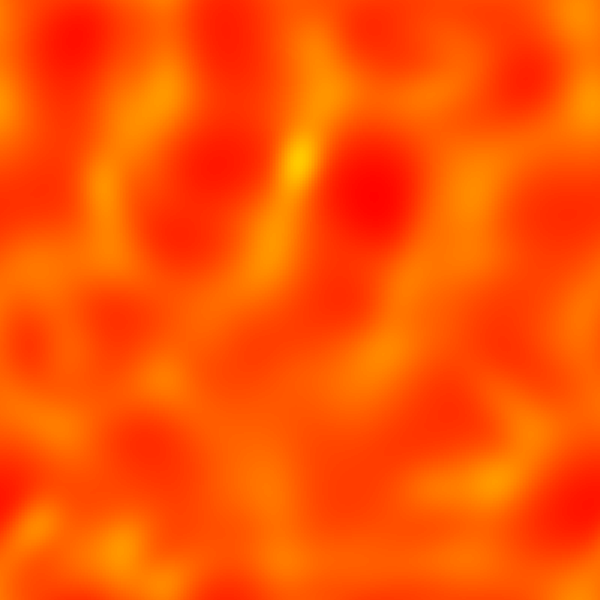}  &
\includegraphics[width=0.15\textwidth]{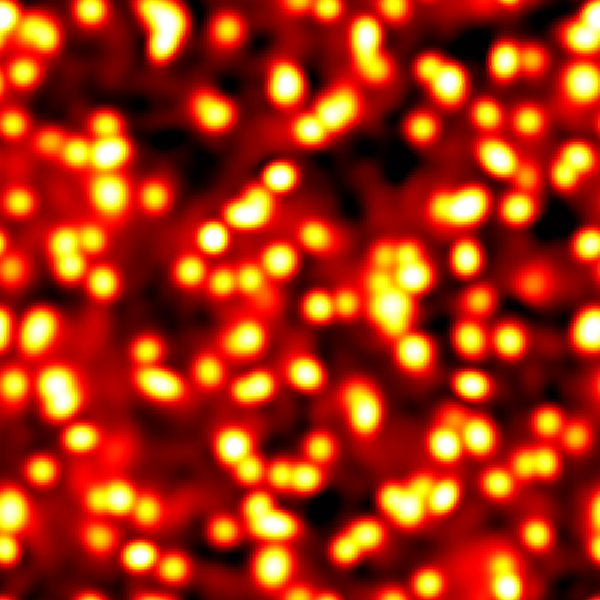} &
\includegraphics[width=0.15\textwidth]{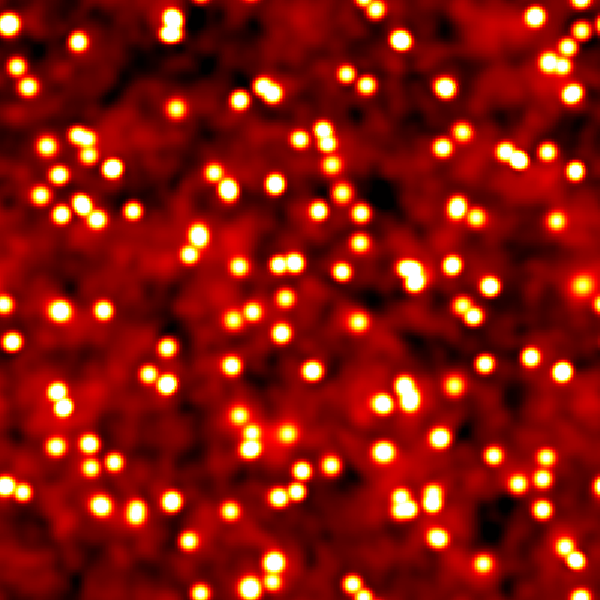} &
\includegraphics[width=0.15\textwidth]{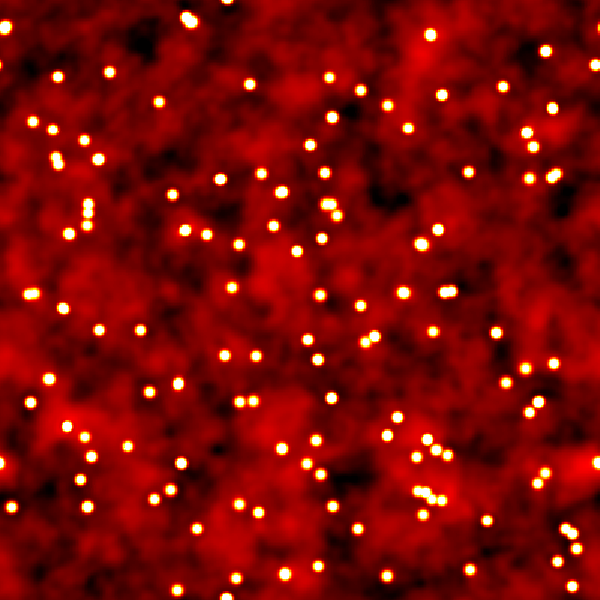} &
\includegraphics[width=0.15\textwidth]{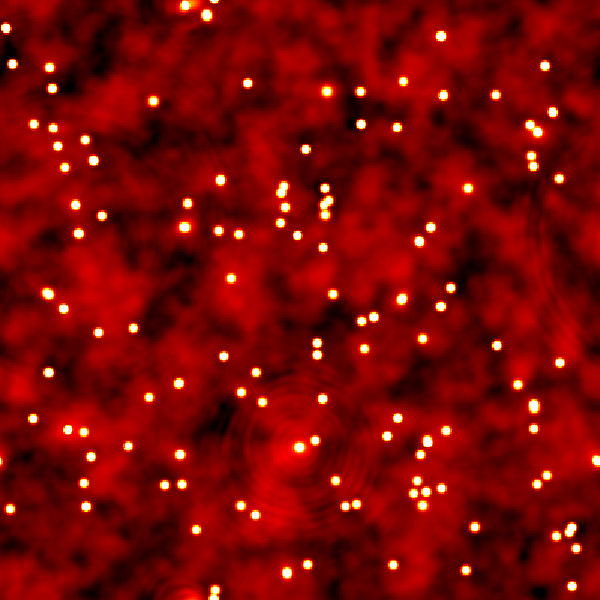} \\
\rotatebox{90}{\,\,\,\,\,\,\,\,\,\,\,no gravity} &
\includegraphics[width=0.15\textwidth]{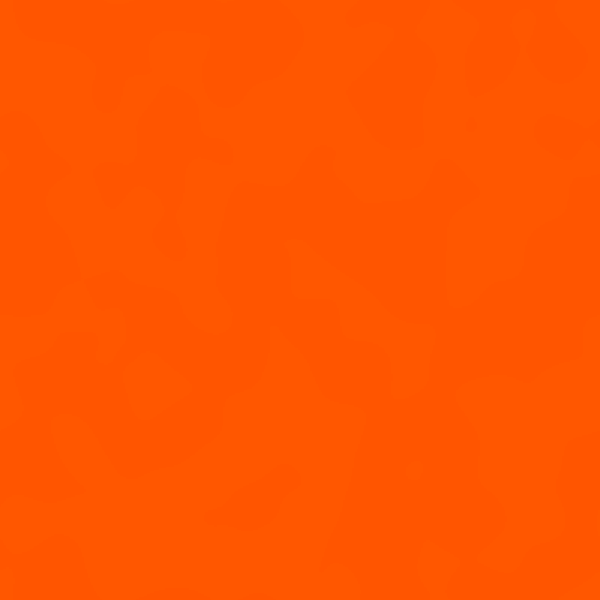}  &
\includegraphics[width=0.15\textwidth]{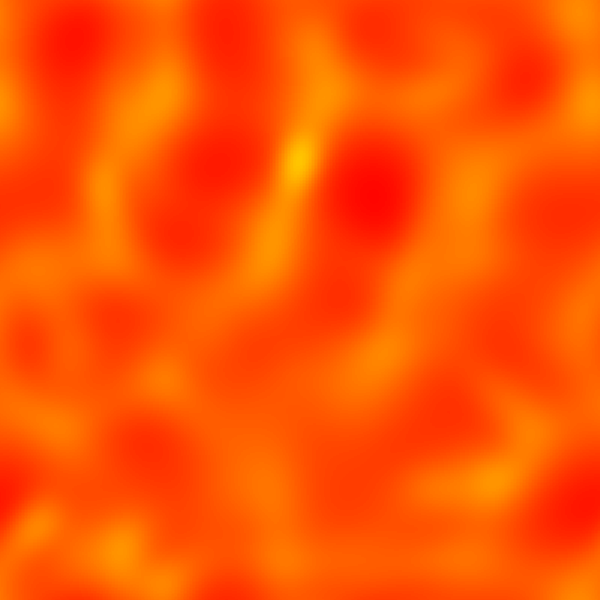}  &
\includegraphics[width=0.15\textwidth]{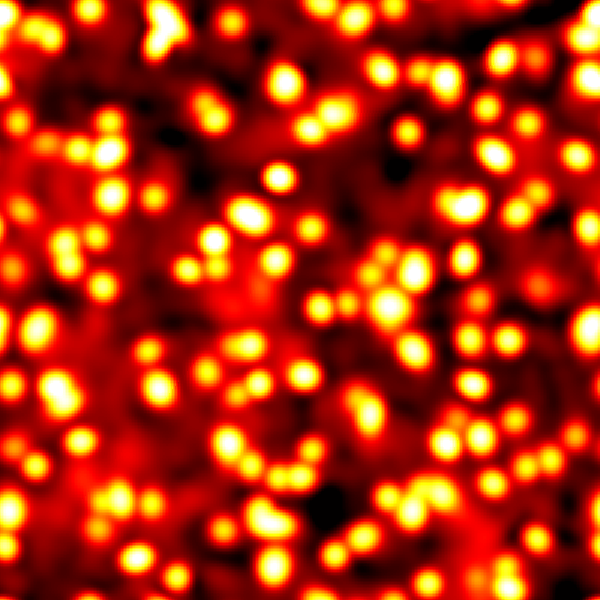} &
\includegraphics[width=0.15\textwidth]{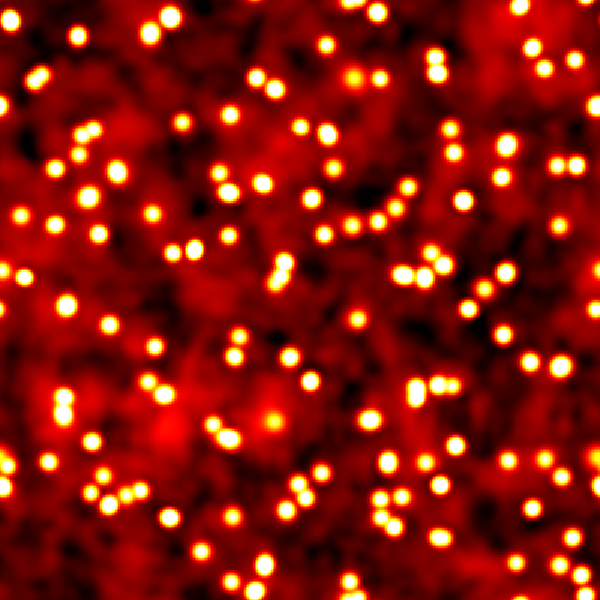} &
\includegraphics[width=0.15\textwidth]{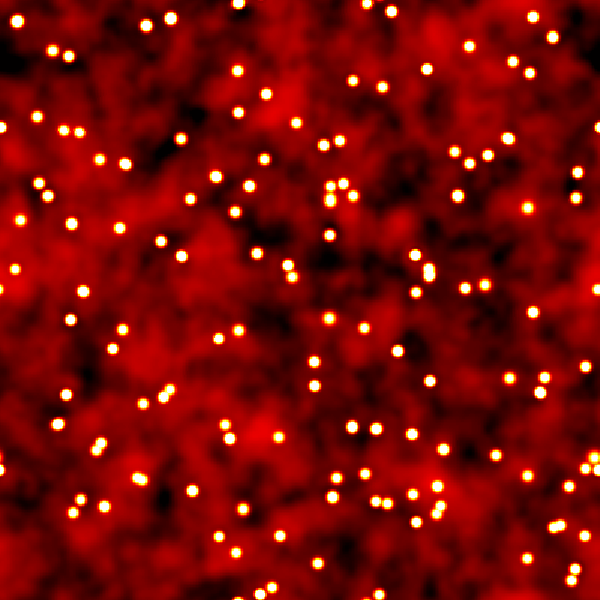} &
\includegraphics[width=0.15\textwidth]{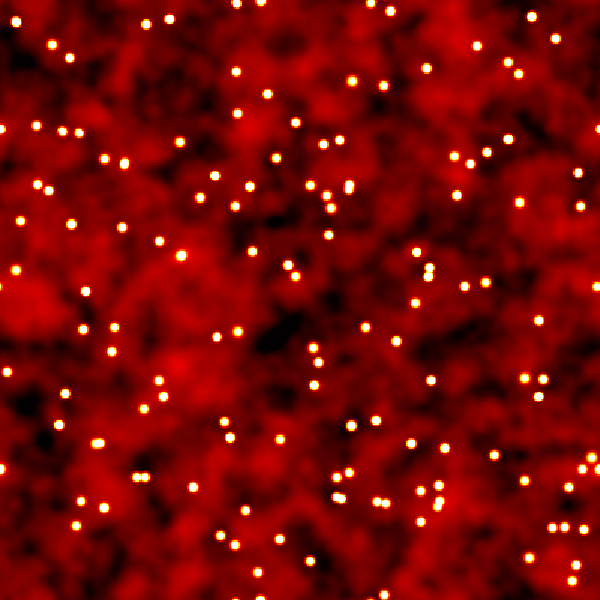} \\
&
\multicolumn{6}{c}{\includegraphics[width=0.25\textwidth]{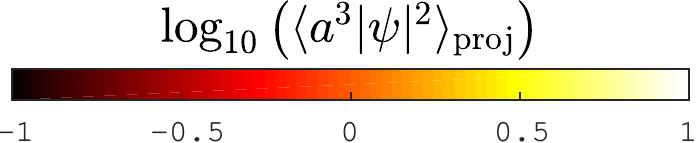}} \\
\end{tabular}
\caption{Projected comoving ``densities'' $a^3|\psi|^2$ (averaged along the line of sight) at several scalefactors ($a=1$ to $a=20$) in our 3+1 dimensional lattice simulations, with $\beta \equiv M/\mpl= 0.03$, and local gravitational interactions switched on (top panels) and off (bottom panels). The early instability due to self-interactions gives rise to the formation of solitons from an almost homogeneous initial state. A statistical analysis of the locations of solitons at late times shows  evidence for clustering only in the case where gravitational interactions are included. Note that inside solitons, $|\psi|^2={\rm const}. $ that is, their core density does not redshift, whereas the background $|\bar{\psi}|^2\propto a^{-3}$.  Moreover, solitons maintain a fixed physical size, hence the illusion of them shrinking in size in a comoving volume. The initial size of the box is the size of the horizon at the beginning of the simulation $L\simeq H_{\rm in}^{-1}$. The solitons contain a dominant fraction of the mass in the simulation volume. On a technical aside, note that the projected comoving density  even in the densest (lightest in color)  regions in the above plot will be smaller than the density inside the cores because of the small volume occupied by the solitons.}
\label{fig:SnapShots}
\end{figure*}

To explore many of these implications, it is important to consider their formation, and their interactions resulting from gravity and self-couplings of the field. In this paper we explore the gravitational clustering and gravitational as well as non-gravitational interactions of non-relativistic solitons, starting with the formation of solitons from  cosmological initial conditions.\footnote{Here, by cosmological initial conditions we mean an almost homogeneous field with small perturbations.} See Fig.~\ref{fig:SnapShots} for a visual overview of soliton formation and clustering in an expanding universe.

We focus on nontopological solitons in a non-relativistic scalar field theory. We include strong self-interactions in the theory, while gravity is included under the assumption that it is weak. In our simulations, the rate of expansion of space is determined by the average energy density of the field. 

There is a large amount of diverse literature on nontopological solitons in real and complex scalar field theories in a cosmological context; this paragraph is a sample rather than a comprehensive review of the literature. For work on individual solitons,  see, for example, \cite{Kaup:1968,Coleman:1985ki,Bogolyubsky:1976nx,Copeland:1995fq,Amin:2010jq,Kasuya:2002zs,Amin:2013ika,2011PhRvD..84d3531C,2011PhRvD..84d3532C,2012A&A...537A.127C,2018PhRvD..98b3009C}. For the early universe, soliton formation in relativistic fields in an expanding universe but ignoring gravitational interactions has been considered, for example, in \cite{Kusenko:1997si,Farhi:2007wj,Amin:2011hj,Gleiser:2011xj}. In the late universe context, gravitational interactions are included in the non-relativistic limit, but self-interactions are ignored or typically assumed to be very weak (e.g.~\cite{Schive:2014dra,2017MNRAS.471.4559M,2016PhRvD..94d3513S}). In this non-relativistic, non-interacting limit, halos and solitons within them have been shown to form.  Binary soliton collisions or interactions and their implications have also been explored under controlled initial conditions (e.g.~\cite{Palenzuela:2006wp,Amin:2014fua,2016PhRvD..94d3513S,Helfer:2018vtq}). The fate of a ``prepared'' collection of relativistic solitons (oscillons) with random velocities was considered in two dimensions and without gravity in \cite{Hindmarsh:2007jb}. The mergers of a small group of pre-existing non-relativistic solitons, with gravity included but without self-interactions, was explored in \cite{2017MNRAS.471.4559M,Schwabe:2016rze}. 

In our work we present the following results for the first time: We simulate and analyze the case of soliton formation with strong self-interactions, starting with cosmological initial conditions. Thereafter, a  ``gas'' of solitons emerges in a self-consistently expanding universe, followed by gravitational clustering of solitons and eventual, dynamically rich, close encounters. We provide a quantitative understanding of the formation, gravitational clustering, individual properties and interactions of solitons based on simulations and analytic calculations. We note that quite generally, we can use the results in the present work to understand the formation and gravitational clustering dynamics of the non-relativistic limits of oscillons, Q-balls, and boson stars with strong self-interactions.\footnote{The emergent, approximate $U(1)$ symmetry in the non-relativistic limit makes  oscillons and Q-balls almost identical \cite{Kasuya:2002zs}, or at the very least obtainable from one another.}

The present work is somewhat related to (but does not rely on) a recent exploration of gravitational perturbations from oscillons and transients \cite{Lozanov:2019ylm}. In \cite{Lozanov:2019ylm}, soliton formation in a relativistic Klein-Gordon equation in an expanding background was investigated; however, gravitational perturbations were calculated passively (i.e.  gravitational clustering was not present). Here, we focus on non-relativistic fields, but clustering due to gravity is included. While the models and context are not identical, a qualitative comparison between relativistic and non-relativistic models and results is discussed in the Appendix.

The rest of the paper is organized as follows. In Section~\ref{sec:Model} we discuss the model for a non-relativistic, self-interacting field in an expanding universe with weak field gravity. In Section~\ref{sec:LS}, we briefly discuss the lattice simulation and our numerical algorithm. The initial conditions for the simulations are provided in Section~\ref{sec:IC}. We analyze linear instabilities from self-interactions and gravitational interactions in Section~\ref{sec:LI}. The numerically calculated power spectrum for the field perturbations is provided in Section~\ref{sec:PS}. In Section~\ref{sec:SF} we discuss the formation of solitons, followed by a discussion of their individual profiles and stability in Section~\ref{sec:IS}.  The gravitational clustering of solitons is discussed in Section~\ref{sec:GC}, and resulting strong soliton interactions are explored in Section~\ref{sec:SSI}. Finally, we present our conclusions and future directions in  Section~\ref{sec:CFD}. In the Appendix we discuss connections to a related relativistic system (at the level of the equations, instabilities, solitons and initial conditions).

\section{The Model}
\label{sec:Model}
We use the following equations of motion (and constraint equations) to explore the dynamics of a non-relativistic, self-interacting, self-gravitating scalar field in an approximately homogeneous and isotropic universe:
\Beq
\label{eq:MasterEq}
&\left[i\left(\partial_t+\frac{3}{2}H\right)+\frac{1}{2a^2}\nabla^2 -U_{\rm nl}'(|\psi|^2) -\Phi\right]\psi=0\,,\\
&\frac{\nabla^2}{a^2}\Phi=\frac{\beta^2}{2}\left[|\psi|^2+\frac{1}{2a^2}|\nabla\psi|^2+U_{\rm nl}(|\psi|^2)\right]-\frac{3}{2}H^2\,,\\
&H^2=\frac{\beta^2}{3}\overline{\left[|\psi|^2+\frac{1}{2a^2}|\nabla\psi|^2+U_{\rm nl}(|\psi|^2)\right]}\,,
\Eeq
where $\overline{[\hdots]}$ indicates a spatial average, $a(t)$ is the scalefactor, $H(t)=\dot{a}(t)/{a}(t)$ is the Hubble rate, $\psi(t,\bx)$ is the complex field amplitude, $\Phi(t,\bx)$ is the Newtonian potential, and $U_{\rm nl}(|\psi|^2)$ encodes the self-interactions of the field.\footnote{We have checked that qualitatively similar results are obtained even if we set $U_{\rm nl}\rightarrow 0$ in the Poisson and Friedmann equations, but keep $U_{\rm n}'(|\psi|^2)\equiv \partial_{|\psi|^2}U_{\rm n}(|\psi|^2)$  in the nonlinear Schr\"odinger equation.} 

All variables and parameters appearing in the above equation are dimensionless. We have expressed time $t$ in units of $\tau_m=\hbar/mc^2$, lengths in units of $\lambda_m=\hbar/mc$, the Newtonian gravitational potential $\Phi$ in units of $c^2$ and $|\psi|^2$ in units of $m^2M^2 c^3/\hbar^3$. Note that $m^2M^2 c^3/\hbar^3$ has dimensions of mass density. We assume that the parameter 
\Beq
\beta \equiv \frac{M}{\mpl}\ll 1\,.
\Eeq

\noindent There are three relevant scales in the equations (not easily discernible in the non-dimensional version): $m=$ mass of particles of our field (without self-interactions), $M$ determines the strength of the self-interactions, and $\mpl$ is the reduced Planck mass which determines the strength of gravity. We work in a parameter regime with: $m\ll M\ll \mpl$. The fiducial value used in the present paper is $M=0.03\,\mpl$ (though we have also varied $M$ by a factor of a few). This particular parameter regime can be natural when identifying $\psi$ as the non-relativistic approximation to the inflaton field \cite{Lozanov:2017hjm} (with $m\simeq 2\times 10^{-4}M$). The hierarchy $m\ll M\ll \mpl$ is also natural for an axion-like field where $M$ plays the role of the decay constant $f$; in this case $m$ can be much smaller (e.g. \cite{Arvanitaki:2009fg,Hlozek:2014lca}). We note that $m$ is essentially setting units of quantities in our equations, and the behavior we explore will be qualitatively valid for any {\it energetically dominant}, cosmological scalar field regardless of the particular value of $m$ (modulo initial conditions).

For the purpose of this paper, we chose $U_{\rm nl}(|\psi|^2)$ with a {\it saturated nonlinearity} :
\Beq
\label{eq:Unl}
U_{\rm nl}(|\psi|^2)=-\frac{|\psi|^2}{2}\frac{|\psi|^2}{1+|\psi|^2}\,.
\Eeq

The saturated nonlinearity refers to the fact that for $|\psi|\gg 1$, $U_{\rm nl}'(|\psi|^2)\rightarrow \rm const.$ which means that the nonlinearity appearing in the equation of motion for $\psi$ is bounded. This form is not strictly necessary, and different powers of $|\psi|^2$ in the denominator of $U_{\rm nl}$ [for example, $(1+|\psi|^2)^\alpha$ or  $(1+|\psi|^{2\alpha})$ with $\alpha>0$], are also worth exploring; however, we do not consider these here.

Note that for $|\psi|^2\ll 1$, the above choice yields $U'(|\psi|^2)=-|\psi|^2$, which makes the first equation in \eqref{eq:MasterEq} analogous to the usual nonlinear Schr\"odinger equation  with attractive interactions (ignoring gravity). Equation \eqref{eq:MasterEq} also matches the equations of motion for axions, or symmetric inflationary potentials in this non-relativistic, small amplitude limit. 

While not necessary for our present purposes, we explore the connection of our non-relativistic equations to those obtained from a relativistic theory in the Appendix. We also refer the reader to (for example) \cite{Namjoo:2017nia,Eby:2018ufi} for more detailed discussions of the non-relativistic limit of relativistic scalar field systems (typically in the weak interaction limit). At the leading order, the non-relativistic limit of real or complex scalar fields should yield equations similar to ours.

\section{Lattice Simulations}
\label{sec:LS}
We solve our Schr\"odinger-Poisson system in a self-consistently expanding background (see eq.~\eqref{eq:MasterEq}) on an $N=400^3$ lattice.\footnote{Smaller lattices were also used to check for convergence, and other numerical checks. For example, we halved the resolution, and the locations of solitons did not change. In our highest resolution simulations, solitons contain $\gtrsim 10$ pixels per linear dimension ($\mathcal{O}[10^3]$ pixels per soliton volume).} The field evolution uses a second-order in time (exponential convergence in space) `kick'-`drift'-`kick' spectral method of \cite{Mocz:2017wlg}.  For our numerical method, the total run time scales as $\mathcal{O}[N^5]$, which limits $N$ from being too large. The initial box size is $L\sim H^{-1}$, and we run our simulations from $a_{\rm i}=1$ to $a_{\rm f}=20$ (with corresponding $t_{\rm f}-t_{\rm i}\simeq {\rm few}\times 10^3\, m^{-1}$). The box size ($L$), resolution ($\Delta x=L/N$), and time duration of the simulations are chosen so that (i) the relevant instability scales (discussed below) are captured in the simulation, (ii) our solitons are resolved ($a_{\rm f}\Delta x\lesssim \mathcal{O}[1]$), and (iii) we have a sufficient number ($\mathcal{O}[10^2]$) of solitons in our simulation volume to make statistically significant statements about their properties, interactions and clustering. 

We find our solitons in the numerical simulations by locating local maxima in the $|\psi|^2$ field (by comparing each pixel to its nearest neighbor in a $3\times 3\times3$ region), and taking all points with comoving density about some threshold.
We look at the radial density profiles about these points and verify that they fall on the central amplitude--radius relation predicted for solitons shown in Fig.~\ref{fig:sol}. In practice, we found that we could distinguish solitons from other local inhomogeneities (which are less dense), with our threshold of $a^3|\psi|^2>25$. The results are invariant to the particular choice of threshold over a range of values: $10$s--$100$s. A lower threshold would start including extraneous linear fluctuations, and a higher threshold would start excluding solitons.


\section{Initial Conditions}
\label{sec:IC}
We begin with an almost homogeneous field with small spatial perturbations (mimicking zero-point fluctuations) of the form
\Beq
\label{eq:IC}
&\psi(t_{\rm in},\bx)=\bar{\psi}(t_{\rm in})+\frac{1}{\sqrt{L^3}}\left(\frac{m}{M}\right)\sum_{\bk} \delta\psi_\bk e^{i\bk\cdot\bx}\,,\\
&{\rm with}\qquad\bar{\psi}(t_{\rm in})=1\qquad{\rm and}\qquad\langle|\delta \psi_\bk|^2\rangle\sim\frac{1}{2}\,.
\Eeq
where $|\delta\psi_\bk|$ are drawn from a Raleigh distribution, and the phases for $\delta\psi_\bk$ are drawn from a uniform distribution.\footnote{Note that $\delta\psi_\bk$ is in units of $m^{1/2}$ (with $\hbar=c=1$). Recall that $\psi$ is measured in units of $mM$ and $L$ in units of $m^{-1}$ which together lead to the appearance of the $m/M\ll 1$ coefficient. To arrive at the above initial conditions, we found it easiest to start from the relativistic case with the relativistic field $\phi=(\sqrt{2}/m)\Re[\psi e^{-imt}]$ (see the Appendix for details). For the initial conditions, we ignore self-interactions, as well as fast time variations and assume $k\lesssim m$. Refinements are possible (such as $|\delta\psi_\bk|^2\sim(1/2)\sqrt{k^2+m_{\rm eff}^2}$), but are not  expected to change the results qualitatively.} We assume that $a(t_{\rm in})=1$. The gravitational perturbations and $H$ are then obtained self-consistently using eq.~\eqref{eq:MasterEq}. We choose $\bar{\psi}(t_{\rm in})=1$ because (as we will see) for $\bar{\psi}(t_{\rm in})\lesssim\beta^{-1}$, instabilities due to self-interactions are ineffective. On the other hand, for $\bar{\psi}(t_{\rm in})\gtrsim \beta^{-1}$ we are forced to introduce a timescale $H^{-1}$ via the Friedmann equation which is comparable to $\tau_m$, thus potentially entering a fast timescale, relativistic regime. 

To remain consistent with our non-relativistic approximation, we introduce a cutoff in the initial spectrum $\langle|\delta \psi_\bk|^2\rangle=0.5e^{-k^2}$ which removes relativistic ($k\gg 1$) modes. We have checked that our results are qualitatively insensitive to order unity changes in amplitudes of the initial perturbations as well as the cutoff.

\section{Linear Instabilities}
\label{sec:LI}
As seen in Fig.~\ref{fig:SnapShots}, there is a rapid growth in field/density perturbations on a characteristic length scale, which results in the formation of solitons. We calculate and compare this instability with gravitational instability below.
\subsection{Self-Interaction Instability}
  Let us consider small spatial perturbations around a homogeneous solution $\bar{\psi}(t)$:
\beq
\label{eq:defPert}
\psi(t,\bx)=\bar{\psi}(t)\left[1+\varepsilon \frac{\delta\psi_\bk(t)}{\bar{\psi}(t)}e^{i\bk\cdot \bx}\right]\,,
\eeq
where $\varepsilon = (m/M)L^{-3/2}$. Sufficiently long wavelength perturbations of the field are unstable due to self-interactions of the field $U'(|\psi|^2)$. To see this, let us first ignore expansion and gravitational interactions (that is, $a=1$, $H=0$, $\Phi=0$), and substitute eq.~\eqref{eq:defPert} into eq.~\eqref{eq:MasterEq}. At the background level, we find $\bar{\psi}(t)=\bar{\psi}(0)e^{-i\nu t}$ with $\nu=U'(|\bar{\psi}|^2)<0$. At linear order in the perturbation, we find\footnote{To obtain this equation, we found it useful to first derive the first-order equations for the real and imaginary parts of the perturbation $e^{i\bk\cdot\bx}\delta\psi_\bk/\psi$ and then combine them to get the second-order-in-time equations for each part. The real and imaginary parts satisfy the same second-order linear equation; thus, we arrive at eq.~\eqref{eq:SIInsEq}.} 
\Beq
\label{eq:SIInsEq}
\left(\partial_t^2+\frac{k^2}{4}\left[k^2+4|\bar{\psi}|^2U_{\rm nl}''(|\bar{\psi}|^2)\right]\right)\frac{\delta\psi_\bk}{\bar{\psi}}=0\,.
\Eeq
Note that $U_{\rm nl}''(|\bar{\psi}|^2)<0$ for our case. Thus, we have unstable, exponentially growing perturbations $|\delta\psi_{\bf k}/\bar{\psi}|\propto e^{\mu_k t}$ for 
\Beq
\label{eq:SI}
&k^2< -4|\bar{\psi}|^2U_{\rm nl}''(|\bar{\psi}|^2)\,,\\
\textrm{with}\quad &\mu_k=\Bigg|i\frac{k}{2}\sqrt{k^2+4|\bar{\psi}|^2U_{\rm nl}''(|\bar{\psi}|^2)}\Bigg|\,.
\Eeq
For a given $|\bar{\psi}|$, the mode that grows the fastest has a wavenumber
\Beq
\label{eq:fastest}
k_\star&= \sqrt{-2|\bar{\psi}|^2U_{\rm nl}''(|\bar{\psi}|^2)}\,\\
\quad\textrm{with}\quad \mu_{k_\star}&=-\bar{\psi}^2U_{\rm nl}''(|\bar{\psi}|^2)=\frac{k_\star^2}{2}\,.
\Eeq
The corresponding (approximate) expressions in an expanding universe, are obtained via  $\bk\rightarrow \bk/a$. Moreover, in an expanding universe $\bar{\psi}\propto a^{-3/2}$ and $H\sim \beta a^{-3/2}$.

In an expanding universe, this growth rate should be compared to $H$ to ascertain whether the growth of perturbations can compete with expansion related dilution. Using our expressions for $U_{\rm nl}(|\psi|^2)$ in eq.~\eqref{eq:SI} and $H^2$ from the Friedman equation \eqref{eq:MasterEq}, we need
\Beq
\frac{\mu_{k}}{H}\sim \frac{1}{\beta}\frac{1}{a^{3/2}}\gg 1\qquad{\rm for}\qquad\textrm{rapid growth}.
\Eeq
In the above expression we have assumed that $|\bar{\psi}|\lesssim 1$.
\subsection{Gravitational Instability}
Spatial perturbations of the field also grow due to gravitational interactions (we ignore self-interactions for the moment). Again, ignoring expansion, usual linear instability analysis of eq.~\eqref{eq:MasterEq} reveals that the unstable perturbations grow exponentially $|\delta\psi_\bk/\bar{\psi}|\sim e^{\mu_k t}$ when \cite{Hu:2000ke}
\Beq
\label{eq:Igrav}
k< k_J\approx \sqrt{\sqrt{2}\beta |\bar{\psi}|}\,\quad{\rm with}\quad \mu_k=\sqrt{\frac{1}{2}\beta^2|\bar{\psi}|^2-\frac{k^4}{4}}\,.
\Eeq
Heuristically, including expansion means that $|\bar\psi|$ and $k$ redshift, and $k$ above should be interpreted as a physical wavenumber $k/a$.\footnote{We recognize that including expansion more carefully, the gravitational instability is power-law type rather than exponential, and the fractional overdensity must grow as $\sim a$ for $k<k_J$; however, our argument is sufficient to capture the slowness of gravitational instability compared to the self-interaction one \cite{Easther:2010mr}.} 
\begin{figure}
\includegraphics[width=3.3in]{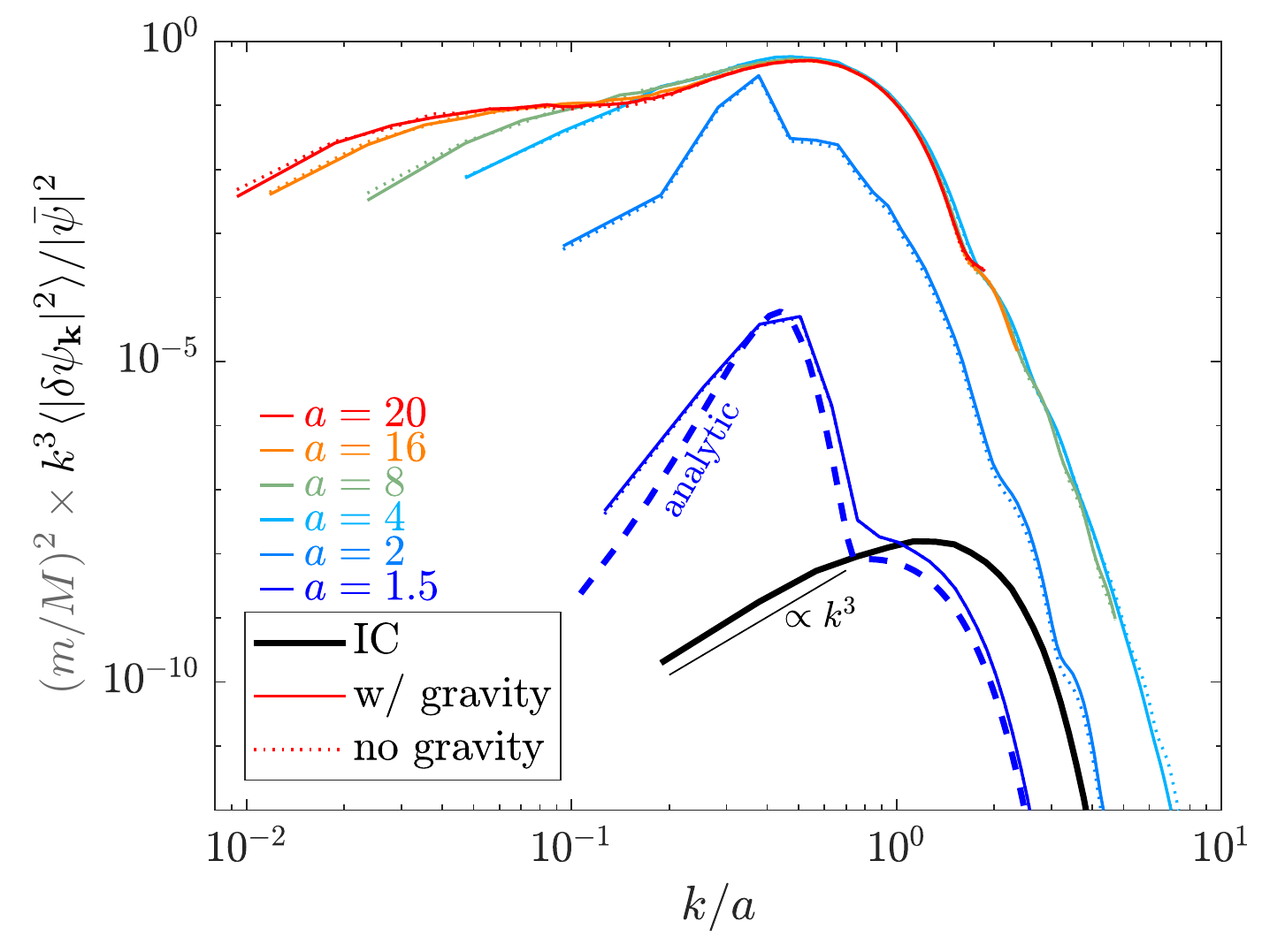}
\caption{Power spectrum of the field $\psi$ (scaled by $|\bar{\psi}|^2\propto a^3$). The initial conditions are consistent with vacuum fluctuations, with a cutoff removing relativistic scales. A self-interaction driven instability on the wavenumber $k/a\approx \sqrt{-2|\bar{\psi}|^2U_{\rm nl}''(|\bar{\psi}|^2)}$ drives the initial growth of the perturbations. These perturbations backreact on the homogeneous condensate around $a_{\rm nl}\simeq 2.1$ on the physical scale $k_{\rm nl}/a\simeq 0.35$ first. After this time, solitons soon begin to form, separated by a comoving distance of  $\sim 2\pi/k_{\rm nl}$. Note that in this figure, since we have divided the power spectrum  $|\bar{\psi}|^2$, the backreaction takes place when the spectrum is roughly of order unity. }
\label{fig:PS}
\end{figure}

We end this section by noting that there are two instability scales associated with self-interactions and gravity respectively (see eqs. \eqref{eq:SI} and \eqref{eq:Igrav}). Assuming $|\bar{\psi}|\sim a^{-3/2}\lesssim 1$, the instabilities are active on physical wave-numbers 
\begin{equation}
\label{eq:comoving}
\frac{k}{a}\lesssim 
\begin{cases}
&a^{-3/2}\qquad \textrm{\qquad self-interactions}\,,\\
&\sqrt{\beta} a^{-3/4} \qquad \textrm{\,\,\,gravity}\,.
\end{cases}
\end{equation}
The unstable modes have characteristic ``growth rates'':
\begin{equation}
\label{eq:ratesCompare}
\frac{\mu_k}{H}\sim
\begin{cases}
&\beta^{-1}a^{-3/2}\qquad \textrm{\,\,\,\,\,self-interactions}\,,\\
&1 \qquad\qquad \qquad\textrm{\,\,gravity}\,.
\end{cases}
\end{equation}
This simple scaling analysis reveals that for $\beta\ll 1$, the self-interaction instability will dominate at early times. 
\section{Power Spectrum}
\label{sec:PS}

The power spectrum of the field perturbations is shown in Fig.~\ref{fig:PS}. The initial spectrum (black) is based on our initial conditions (see eq.~\eqref{eq:IC}, including an exponential cutoff which removes $k\gg 1$ modes at this time). 

The dashed blue line is the expected power spectrum at $a=1.5$ based on our instability analysis in  Section~\ref{sec:LI}. This calculated power spectrum is consistent with the numerically evaluated spectrum at the same time which was obtained using the full lattice simulation, both with local gravitational interaction included (solid line) and turned off (dotted line).\footnote{We note that there is some power on $k/a\gtrsim 1$ in the power spectrum; part of this is from initial conditions where we were not aggressive in removing all $k/a\gtrsim 1$ modes, and part from rescattering due to nonlinear evolution. However at late times, most of the power is on $k/a\lesssim 1$.} 

Soon after, the perturbations start becoming nonlinear, and backreaction of the perturbations on the homogeneous evolution of the field becomes significant. The scalefactor when the perturbations become nonlinear can be obtained from the following heuristic criterion which compares the amplitude of field perturbations to the background homogeneous field:
\Beq
\label{eq:backreaction}
\frac{m}{M}k^{3/2}\langle |\delta\psi_\bk|^2\rangle^{1/2} \sim \bar{\psi}\,,\\
\Eeq
where the left-hand side is an estimate of the variance of fluctuations on a scale $l\sim k^{-1}$. The above criterion is satisfied by a combination $(a_{\rm nl},k_{\rm nl})$ such that the field perturbations on the comoving scale $k_{\rm nl}$ become nonlinear first. For $\beta=0.03$, we analytically estimate $a_{\rm nl}\simeq 2.1$ and $k_{\rm nl}\simeq  0.7$. Note this scale  $k_{\rm nl}/a\simeq 0.35$ in the spectrum in Fig~\ref{fig:PS} (see the blue curves). A characteristic scale is also visible in the second column ($a=2$) of the snapshots of the field evolution shown in Fig.~\ref{fig:SnapShots}. 
\begin{figure}
\includegraphics[width=3.4in]{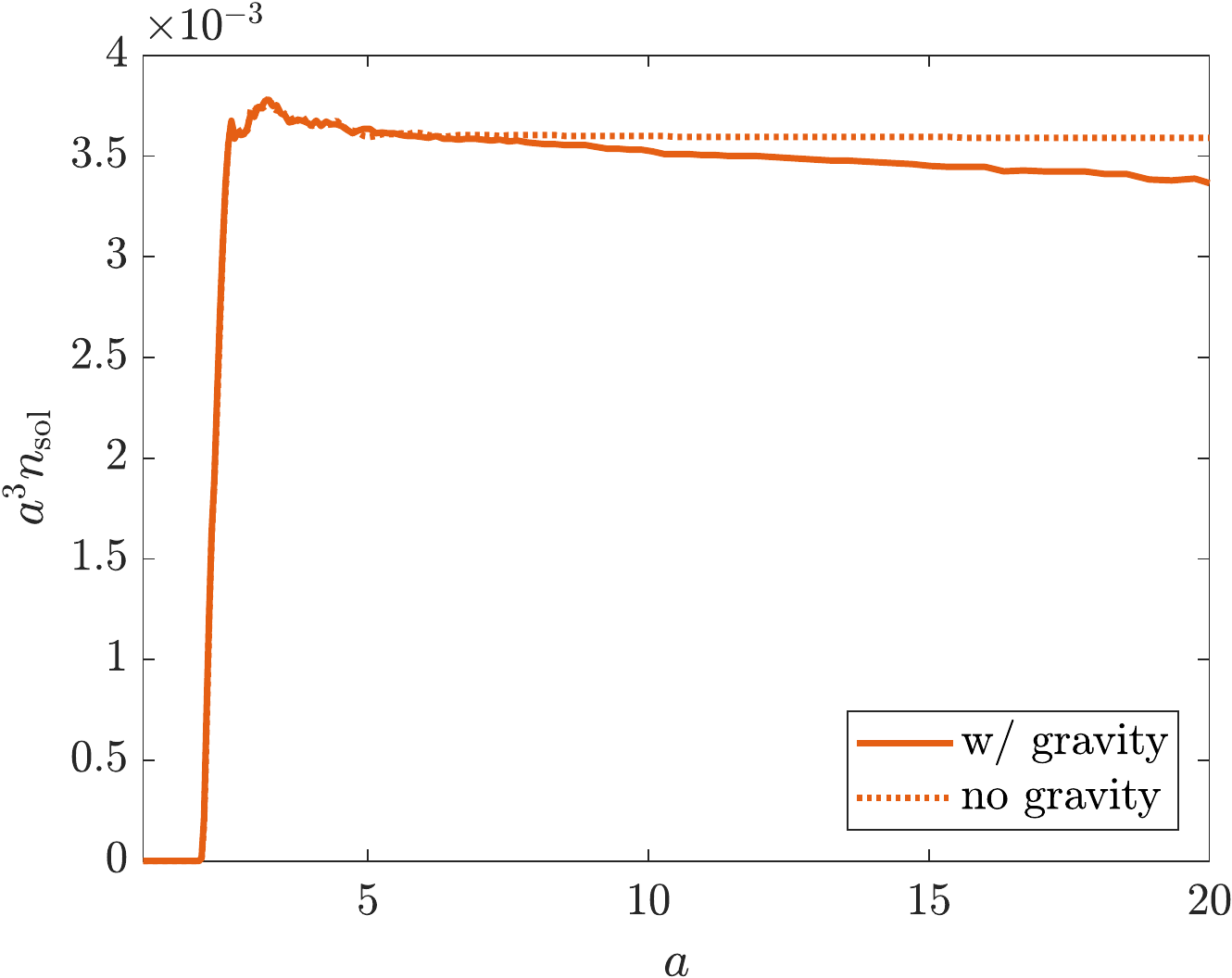}  
\caption{
The figure shows the comoving number density of solitons $a^3n_{\rm sol}$ in our simulations with (solid) and without (dotted) gravitational interactions. Proper solitons begin to form around $a\approx 4$, with $\mathcal{O}[10^3]$ solitons per Hubble volume $H^{-3}$ at this time. At late times, the number density of solitons is lower in the case when gravity is included due to mergers and disruptions made possible by gravitational clustering. The curves are obtained by averaging over 6 runs.}
\label{fig:ns}
\end{figure}
\section{Soliton Formation}
\label{sec:SF}
Once the perturbations become nonlinear, the attractive self-interactions lead to the formation of localized, roughly spherical energy density configurations (our solitons) at the peaks of the density perturbations. The comoving number density of such peaks (and hence of solitons) is crudely given by
\Beq
a^3n_{\rm sol}\sim (k_{\rm nl}/2\pi)^3\,,
\Eeq
at the time of formation (see \cite{Amin:2010xe,Amin:2010dc}). Using $k_{\rm nl}\simeq  0.7$,  we get $a^3n_{\rm sol}\sim 10^{-3}$, consistent with our simulations (see Fig.~\ref{fig:ns} ). 

The formation of solitons following the initial linear instability is clearly visible in the snapshots shown in Fig.~\ref{fig:SnapShots}. While we do not show the $a=3$ snapshot, the formation of solitons is complete by this time. The $a=4$ snapshot shows well-formed, and separated solitons, with typical overdensity inside the solitons of $\mathcal{O}[10]$.

In more detail, Fig.~\ref{fig:ns} shows the comoving number density of solitons as a function of time in our simulations. The initial number density established by the formation of the solitons is independent of self-gravity. However, gravity is strong enough to lead to subsequent mergers or disruptions which leads to a small drop in number density of solitons at late times.  In addition, we cannot rule out that gravity is causing some individual solitons to become unstable. The drop in comoving number density is evident in the difference between the dashed (ignoring gravitational interactions) and solid lines ($\lesssim 10\%$ per Hubble time).

We find that a large fraction ($\sim 70\%$) of the energy in a comoving volume of the universe is locked up in solitons. We only count regions with over-densities $\gtrsim 4$ as part of solitons for this estimate. This result is consistent with related earlier simulations using the relativistic nonlinear Klein-Gordon equation in an expanding universe (but ignoring gravitational clustering); see for example \cite{Amin:2010dc,Amin:2011hj}.

\section{Individual Solitons}
\label{sec:IS}
The first two equations in eq.~\eqref{eq:MasterEq} (ignoring expansion) admit spatially localized, spherically symmetric, solitonic solutions of the form
\Beq
\psi(t,r)=e^{-i\nu t}\Psi(r)\,.
\Eeq
We substitute this ansatz into \eqref{eq:MasterEq} to obtain equations for the profile $\Psi(r)$ and gravitational potential $\Phi(r)$:
\Beq
\label{eq:profiles}
&\left[\nu+\frac{1}{2r^2}\partial_r(r^2\partial_r)-U_{\rm nl}'(\Psi^2) -\Phi\right]\Psi=0\,,\\
&\frac{1}{r^2}\partial_r(r^2\partial_r)\Phi=\frac{\beta^2}{2}\left[\Psi^2+\frac{1}{2}(\partial_r\Psi)^2+U_{\rm nl}(\Psi^2)\right]\,.
\Eeq
Note that $\nu$ can be absorbed into the definition $\tilde{\Phi}=\Phi-\nu$. We then find smooth, localized, node-free solutions for $\Psi(r)$ for each $\Psi(0)$, by appropriately adjusting $\tilde{\Phi}(0)$.\footnote{If needed, we can recover $\nu=\Phi-\tilde{\Phi}$ by insisting that $\Phi(r)\rightarrow 0$ for $r\rightarrow \infty$. In practice, recovering accurate values of $\nu$ is not easy since $\tilde{\Phi}$ falls off as a power law.}

We note that by going to the large $r$ limit of the profile equations, $\Psi(r)$ decays in an exponential fashion at large radii (see \cite{Schiappacasse:2017ham}). This will be relevant when discussing soliton interactions. 

In Fig. \ref{fig:sol} we plot the $1/e$ width of these soliton profiles as a function of the central amplitude (solid black curve) using the profiles obtained from the above procedure. Note that the width is non-monotonic in the central amplitude. The data points in this plot correspond to solitons extracted from our simulations and are in excellent agreement with the calculated analytic expectation. Note that for early times ($a= 2$), not all high density regions are solitons yet; hence, they do not lie on the analytic curve initially.

While we have done the above calculation including gravity, the gravitational potential remains small for most of the solitons: $|\Phi(0)|=\mathcal{O}[10^{-3}]$ for $\beta= \mathcal{O}[10^{-2}]$, and gravity does not significantly affect profiles for central amplitudes $\Psi(0)\lesssim \rm few$. The same is true in our simulations. We also show the gravitational potential at the center of these solitons Fig.~\ref{fig:sol} (top axis). 

The mass (or energy) per soliton is\footnote{Note that ignoring the gradient and potential terms only changes the answer by a factor of few. We briefly restore units with $\hbar=c=1$ to clarify that each soliton contains a large number of $m$ particles.}
\Beq
\mathcal{E}&=\int d^3 r \left[\Psi^2+\frac{1}{2}(\partial_r\Psi)^2+U_{\rm nl}(\Psi^2)\right]\,,\\
&=\mathcal{O}[10^2]\times\left(\frac{M}{m}\right)^{\!\!2}m\,,
\Eeq
for the range of central amplitudes shown in Fig.~\ref{fig:sol} and seen in simulations. Note that with $m\ll M$, $\mathcal{E}\gg m$. We find that the energy is a non-monotonic function of $\Psi(0)$, with a minimum near $\Psi(0)\simeq 1$. \subsection*{Stability}
 From our calculated profiles, we find that for $-\nu\gtrsim 0.05$ (correspondingly, $\Psi(0)\gtrsim 0.9$):
\Beq
\frac{d\mathcal{N}}{d(-\nu)} > 0\qquad \textrm{where}\qquad \mathcal{N}\equiv \int d^3r \Psi^2(r)\,,
\Eeq
whereas it is smaller than zero at smaller amplitudes. This Vakhitov-Kolokolov stability criterion \cite{Vakhitov:1973} guarantees stability for solitons with $\Psi(0)\gtrsim 0.9$ against long-wavelength perturbations.

\begin{figure}
\includegraphics[width=3.3in]{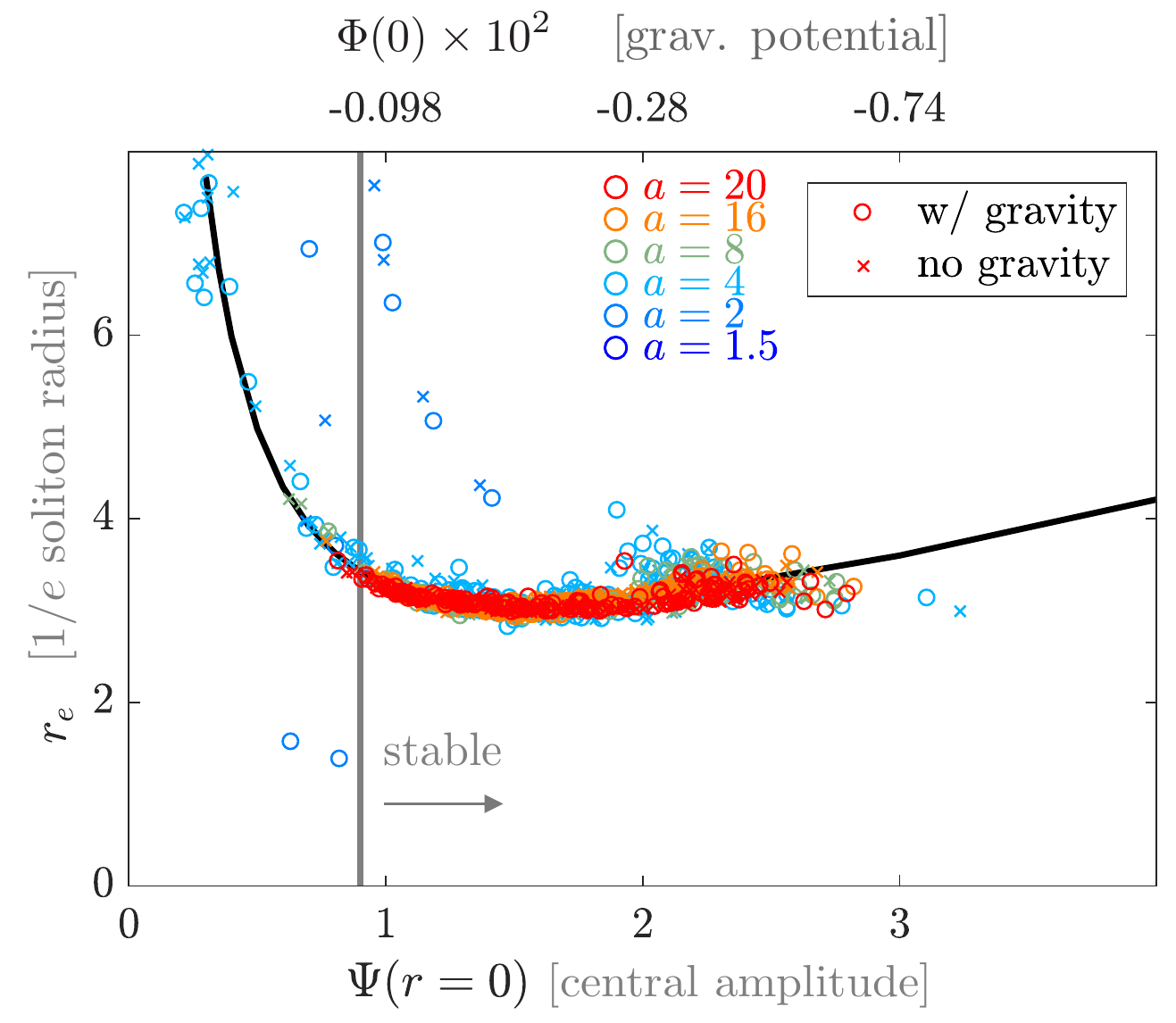}  
\caption{
The relationship between the central amplitude and $1/e$ width of the solitons. The points are extracted from our simulations, whereas the curve is calculated semi-analytically. Note that at late times, only solitons that are stable according to the Vakhitov-Kolokolov stability criterion (on the right of the gray line) remain. For our parameters, gravity remains weak  and does not  significantly alter individual soliton profiles. The gravitational potential at the center of the solitons is plotted on the top axis.}
\label{fig:sol}
\end{figure}

The stability criterion elegantly explains the dearth of solitons with central amplitudes below $\Psi(0)\lesssim 1$ in Fig.~\ref{fig:sol}(see also \cite{Eby:2018ufi}, where this criterion is argued to hold even when including gravity in the non-relativistic limit).\footnote{A long-wavelength stability analysis for relativistic solitons (oscillons) was carried out in \cite{Amin:2010dc,Amin:2010jq} (albeit in a different self-interaction potential, and without gravity), which also showed that the above stability criterion correctly predicted the survival of large amplitude oscillons in simulations. We further note that three-dimensional oscillons in sine-Gordon potentials (for axions, but without gravity) are not stable and have a relatively short lifetime, compared to flattened potentials \cite{Salmi:2012ta,Amin:2011hj}.  Oscillons in flattened potentials can last longer than $10^7 m^{-1}$\cite{Salmi:2012ta}, whereas the duration of our simulations is $t_{\rm f}-t_{\rm i}\sim {\rm few} \times10^3 m^{-1}$. See the Appendix for further references on lifetimes in the relativistic case.} A more detailed stability analysis including gravity for our saturated potentials would be useful.\footnote{For a related analysis in the case of  axions, see \cite{Hertzberg:2010yz,Visinelli:2017ooc}.} 
\section{Gravitational Clustering}
\label{sec:GC}
For $\beta\ll 1$, gravitational clustering is expected to become important at late times (significantly after the solitons have formed, see eq.~\eqref{eq:ratesCompare}). At these late times, this universe essentially behaves as a matter dominated universe ($a(t)\propto t^{2/3}$), with solitons becoming our new non-relativistic dust particles on scales much larger than their size. As a result, our zeroth order expectation is that the gravitational clustering of these solitons should proceed in a manner similar to dust in an expanding universe. Moreover, we can ignore non-gravitational forces between the solitons at separations much larger than $2r_e$ because we expect them to be Yukawa-like, with the force falling away exponentially with separation.\footnote{This is also reminiscent of the force between solitons as analyzed by \cite{Manton:1978gf}.}

\begin{figure}
\includegraphics[width=3.4in]{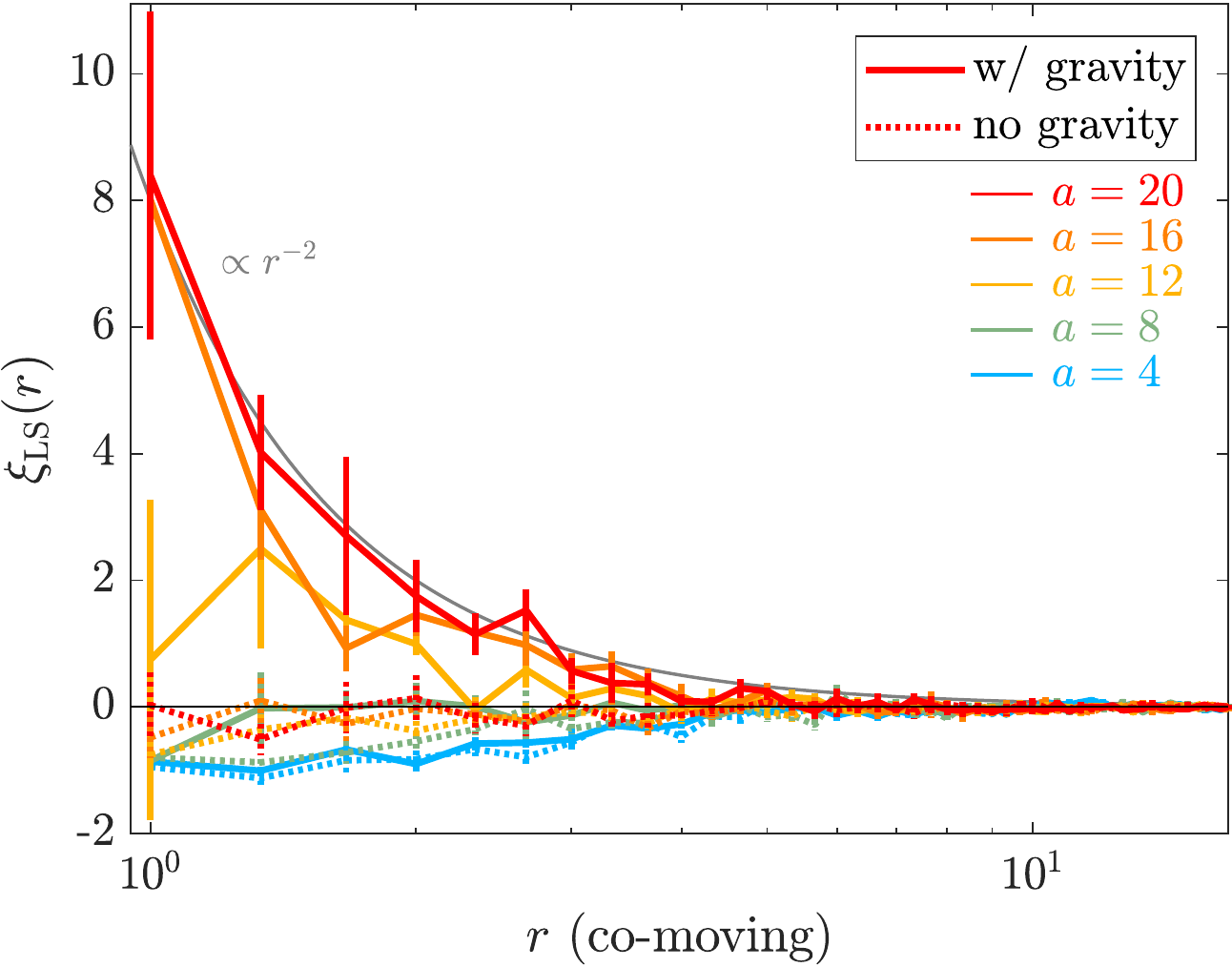}  \caption{The two point correlation function of soliton locations with and without the inclusion of gravitational interactions. At early times, the correlation function with and without gravity agree with each other. However, at late times gravitational clustering $\xi_{\rm LS}(r)\propto r^{-2}$ is clearly visible for the $a=16$ and $a=20$ cases in the above figure.}\label{fig:zeta}
\end{figure}

We construct the two-point correlation function of {\it soliton locations} obtained from our simulations to quantitatively investigate the effects of gravitational clustering. In Fig.~\ref{fig:zeta}, we show the two-point correlation function of the solitons, calculated with the Landy-Szalay estimator \cite{1993ApJ...412...64L,2012psa..book.....W}:
\begin{equation}
\xi_{\rm LS}(r)= \frac{DD}{RR} - 2\frac{\mathcal{N}-1}{\mathcal{N}}\frac{DR}{RR} + 1\,,
\end{equation}
where there are $\mathcal{N}$ solitons (the data $D$), and $\mathcal{N}$ uniform randomly chosen points $R$, and $DD$ is the number of soliton pairs in a given comoving radial separation bin, $RR$ is the mean count for the random points over several realizations $R$, and $DR$ is the cross-correlation statistic.

As seen in  Fig.~\ref{fig:zeta}, the measured two point correlation function is the same for the cases with and without gravitational interactions at early times soon after soliton formation ($a\lesssim 4$). The distribution is close to Poissonian on large scales: $\xi_{\rm LS}(r\gtrsim 10)\approx 0$. However, the comoving scale $r_{\rm nl}\sim k_{\rm nl}^{-1}$ which is the typical separation of solitons when they first form manifests itself in a negative correlation function on small scales (we find very few solitons with separations less than $k_{\rm nl}^{-1}$).

If we allow for gravitational interactions, solitons begin to cluster. This clustering can be quantified in our simulations at late times as excess power in $\xi_{\rm LS}$ (for $a\gtrsim 10$). Consistent with clustering of point particles in a matter dominated universe starting with uncorrelated positions \cite{1980ApJ...235..299S}, we find 
\beq
\xi_{\rm LS}(r)\propto \frac{1}{r^2}\,,
\eeq
where $r$ is a comoving separation.\footnote{We checked that if we replace the solitons by point particles after $a=4$, the correlation function evolves in a qualitatively similar manner} Fitting the model $\xi_{\rm LS}\propto a^\alpha r^{\beta}$ for our 6 simulations in the range of $a=10$ to $a=20$, we find $\alpha = 1.7\pm 0.3$, $\beta = -2.1\pm 0.2$. It would be interesting to explore this clustering further in detail, since it might reveal differences from the point particle case at late times.

\begin{figure*}[t!]
\includegraphics[width=0.7\textwidth]{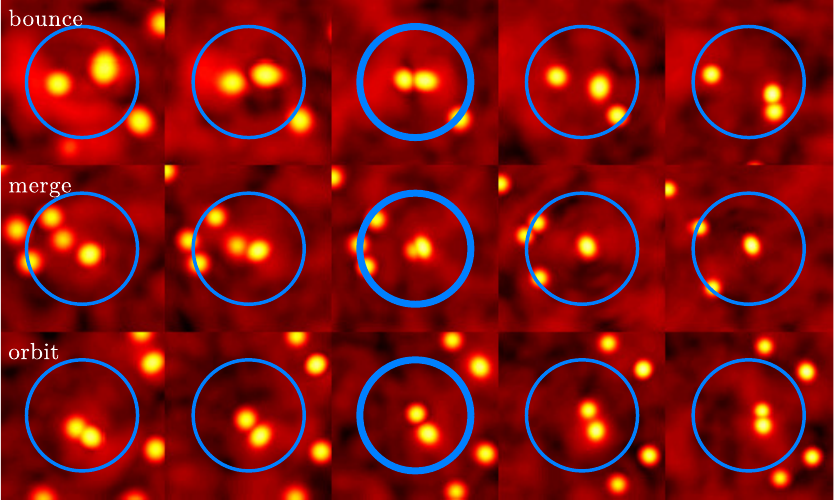}
\caption{Gravitational clustering facilitates close encounters at late times between solitons. Such close encounters lead to mergers, strong scattering and formation of soliton binaries. non-gravitational interactions can play a dominant role in the close encounters, with the phase of the scalar field also playing an important role. This richness in the close-encounter dynamics makes the soliton gas distinct from a gravitationally interacting gas of particles. Shown in this figure are projected densities in zoom-ins (box size $L/4$), around 3 interactions (bounce, merge, and orbit), at 5 times, each separated by time interval corresponding to $\Delta\log(a)=1.16$.}
\label{fig:Interactions}
\end{figure*}
\section{Strong Soliton Interactions}\label{sec:SSI}
Self-gravity plays the important role of bringing solitons together at late times (i.e., significantly after their formation), and allows them to interact.\footnote{There are interactions at early times when gravity is ignored as well, but this is not so at late times in our simulations. We find that the some solitons have a significant velocity at early times with and without gravity, which will be investigated quantitatively in the future.} Figure~\ref{fig:Interactions} shows three different types of interactions that are achieved from our cosmological initial conditions. \begin{enumerate}
\item Solitons ``repel or bounce off'' each other when the relative phase of the interacting solitons $|\theta_1-\theta_2|\approx \pi$ where $\psi_{a}(t,\bx)=\Psi_a(\bx)e^{-i(\nu_{a} t+\theta_{a})}$ with $a=1,2$. We have verified this phase structure in our simulations during such a repulsive interaction. 

\item A few solitons merge to form more massive solitons (typically when the relative phase is $\sim 0$), resulting in a change in the number density of solitons. Such interactions are typically accompanied by the generation of a burst of scalar waves as the solitons settle into new configurations. 
\item A small fraction of solitons form orbiting binaries, and we even see an occasional three-body interaction.
\item Only ${\rm few}-10\%$ of the number of solitons in our simulations undergo strong encounters per Hubble time.\footnote{We inspected 6 numerical runs with different initial conditions to get this number.} This is consistent with the rate of change in the comoving number density of solitons 
\beq
\frac{d\ln (a^3n_{\rm sol})}{d\ln a} \simeq 0.1\,,
\eeq
as seen from Fig. \ref{fig:ns}. 
\end{enumerate}

We re-iterate that bouncing, binary formation, and merging of solitons are self-consistently obtained from our cosmological initial conditions. Evidently, the dynamics of these strong interactions are quite rich, and deviate from the expectations of treating these solitons as just point particles. The relative phase of the solitons plays an important role in these close encounters. 

We note that at late times ($a_{\rm f}\simeq 20$), we have about 10 pixels per linear dimension of the soliton ($\sim 10^3$ pixels per volume of the soliton). As a result,  the detailed dynamics (such as post-interaction kicks at late times) of individual strong interactions should be interpreted with some care. While it is not easy to improve the resolution significantly for the entire simulation, zoom-in, higher resolution simulations focusing on soliton interactions using initial conditions from our simulations would be useful. A more detailed investigation of the rich dynamics of close encounters with higher resolution simulations is left for future work. For an early, and detailed investigation of $Q$-ball interactions ({\it relativistic} complex field valued analogs of our solitons), but without gravity, see \cite{Axenides:1999hs,Battye:2000qj}. 

The repulsive and attractive behavior of such solitons as a function of relative phase can be heuristically understood as follows. Consider a probe soliton moving past another stationary soliton (in the absence of gravity). The nonlinearity in the Schr\"{o}dinger equation ($\propto |\psi|^2$ for $|\psi|^2\ll 1$) can be thought of as a nonlinear refractive index.\footnote{This is more than an analogy since nonlinear Schr\"{o}dinger equations are used to model light pulse propagation in nonlinear media  \cite{Aitchison:91}; we learned of the above heuristic explanation from the same paper. For soliton formation and interactions in yet another context (Bose-Einstein condensates), see for example \cite{2017Sci...356..422N}.} If the two solitons are in phase, we expect this term to be larger in the region between the solitons than the case when the stationary soliton is absent. It also increases towards the stationary soliton. As a result, this larger refractive index, and its gradient, will cause the core of the probe soliton to bend towards the stationary one; i.e. there will be attraction between the solitons. On the other hand, when our two solitons are out of phase, the $|\psi|^2$ between the two solitons will be smaller and have to go to zero in the middle (from symmetry), causing the probe soliton to move away from the stationary one (hence ``repulsion''). A more detailed, effective potential based analysis at large separations is provided by \cite{Palenzuela:2006wp,Cotner:2016aaq}.

\section{Conclusions \& Future Directions}
\label{sec:CFD}
We investigated the dynamics of non-relativistic scalar fields in an expanding background. By including self-interactions and gravitational interactions, we demonstrated the formation of solitons driven by self-interactions from cosmologically relevant initial conditions, followed by gravitational clustering of solitons. We showed that this clustering leads to dynamically rich interactions between solitons including scattering, merging and binary formation at late times (which is absent in the case when gravity is not included). The highly nonlinear dynamics were explored by numerically solving the Schr\"{o}dinger-Poisson system of equations with self-interactions and weak field gravity in a self-consistently expanding universe.  

We provided analytic results and estimates for (i) the time scales and length scales associated with soliton formation, (ii) the spatial distribution of solitons, (iii) the number density of solitons, (iv) the individual properties of our three-dimensional solitons, including their stability, and (v) the two-point function related to the gravitational clustering of solitons.

We showed agreement between our analytic calculations and numerical simulations. The estimates and analytic results also provide an understanding of how the results depend on essential physical parameters in our problem, allowing for broader applicability beyond that of the fiducial models considered in this paper. In the Appendix we discuss the connection of our work to the case where the fields satisfy a relativistic Klein-Gordon equation in an expanding universe (in particular, \cite{Lozanov:2019ylm}). A more careful comparison with relativistic simulations, and many associated subtleties and caveats, is left for future work.

Our work points towards a number of new avenues of exploration: (1) What is the end state of a gravitationally and non-gravitationally interacting ``soliton gas''? What is the velocity and angular momentum distribution?  This investigation is not purely gravitational because of the close encounters of the solitons in an expanding universe, where the phase plays a dominant role (see \cite{Schwabe:2016rze} for the non-interacting case). (2) For our initial conditions,  individual solitons seem to be far from forming black holes. However, rare, accidental over-densities or over-densities driven by gravitational clustering and mergers might make it more favorable to form black holes. Numerically intensive, general relativistic simulations of soliton formation from cosmological initial conditions and strong self-interactions have not yet been done \cite{Helfer:2016ljl,Widdicombe:2018oeo}. (3) The close encounters could be a source of stochastic gravitational waves from solitons in the early universe, in addition to those from formation of the solitons in the early universe. (4) It is possible to consider a different expansion history of the background (for example, radiation domination) and an axion-like potential as well as inhomogeneous initial conditions, which would make parts of our analysis relevant for the formation of quasi-stable axitons \cite{Kolb:1993hw} and axion miniclusters \cite{Hogan:1988mp} in the early universe.\footnote{Radiation domination makes soliton formation and clustering more difficult starting from approximately homogeneous initial conditions.} 
\\

\section*{Acknowledgments} We thank Kaloian D. Lozanov for help in comparing the results in this paper to relativistic simulations in an expanding background and for helpful comments on the manuscript, Marcos G. Garcia for useful discussions regarding the connection between non-relativistic/relativistic field theories, Rohith Karur for checking the importance of gravitational interactions in soliton profiles, and Eugene E. Lim for helpful comments on the draft. Part of the simulations were carried out on the Odyssey cluster supported by the FAS Division of Science, Research Computing Group at Harvard University. MA is supported by a DOE grant DE-SC0018216. Support (PM) for this work was provided by NASA through Einstein Postdoctoral Fellowship grant number PF7-180164 awarded by the \textit{Chandra} X-ray Center, which is operated by the Smithsonian Astrophysical Observatory for NASA under contract NAS8-03060. Part of this work was carried out at the Aspen Center for Physics, which is supported by National Science Foundation grant PHY-1607611.

\bibliography{mybib}{}

\begin{thebibliography}{77}%
\makeatletter
\providecommand \@ifxundefined [1]{%
 \@ifx{#1\undefined}
}%
\providecommand \@ifnum [1]{%
 \ifnum #1\expandafter \@firstoftwo
 \else \expandafter \@secondoftwo
 \fi
}%
\providecommand \@ifx [1]{%
 \ifx #1\expandafter \@firstoftwo
 \else \expandafter \@secondoftwo
 \fi
}%
\providecommand \natexlab [1]{#1}%
\providecommand \enquote  [1]{``#1''}%
\providecommand \bibnamefont  [1]{#1}%
\providecommand \bibfnamefont [1]{#1}%
\providecommand \citenamefont [1]{#1}%
\providecommand \href@noop [0]{\@secondoftwo}%
\providecommand \href [0]{\begingroup \@sanitize@url \@href}%
\providecommand \@href[1]{\@@startlink{#1}\@@href}%
\providecommand \@@href[1]{\endgroup#1\@@endlink}%
\providecommand \@sanitize@url [0]{\catcode `\\12\catcode `\$12\catcode
  `\&12\catcode `\#12\catcode `\^12\catcode `\_12\catcode `\%12\relax}%
\providecommand \@@startlink[1]{}%
\providecommand \@@endlink[0]{}%
\providecommand \url  [0]{\begingroup\@sanitize@url \@url }%
\providecommand \@url [1]{\endgroup\@href {#1}{\urlprefix }}%
\providecommand \urlprefix  [0]{URL }%
\providecommand \Eprint [0]{\href }%
\providecommand \doibase [0]{http://dx.doi.org/}%
\providecommand \selectlanguage [0]{\@gobble}%
\providecommand \bibinfo  [0]{\@secondoftwo}%
\providecommand \bibfield  [0]{\@secondoftwo}%
\providecommand \translation [1]{[#1]}%
\providecommand \BibitemOpen [0]{}%
\providecommand \bibitemStop [0]{}%
\providecommand \bibitemNoStop [0]{.\EOS\space}%
\providecommand \EOS [0]{\spacefactor3000\relax}%
\providecommand \BibitemShut  [1]{\csname bibitem#1\endcsname}%
\let\auto@bib@innerbib\@empty
\bibitem [{\citenamefont {Lee}\ and\ \citenamefont {Pang}(1992)}]{Lee:1991ax}%
  \BibitemOpen
  \bibfield  {author} {\bibinfo {author} {\bibfnamefont {T.~D.}\ \bibnamefont
  {Lee}}\ and\ \bibinfo {author} {\bibfnamefont {Y.}~\bibnamefont {Pang}},\
  }\href {\doibase 10.1016/0370-1573(92)90064-7} {\bibfield  {journal}
  {\bibinfo  {journal} {Phys. Rept.}\ }\textbf {\bibinfo {volume} {221}},\
  \bibinfo {pages} {251} (\bibinfo {year} {1992})},\ \bibinfo {note}
  {[169(1991)]}\BibitemShut {NoStop}%
\bibitem [{\citenamefont {Vilenkin}\ and\ \citenamefont
  {Shellard}(2000)}]{Vilenkin:2000jqa}%
  \BibitemOpen
  \bibfield  {author} {\bibinfo {author} {\bibfnamefont {A.}~\bibnamefont
  {Vilenkin}}\ and\ \bibinfo {author} {\bibfnamefont {E.~P.~S.}\ \bibnamefont
  {Shellard}},\ }\href
  {http://www.cambridge.org/mw/academic/subjects/physics/theoretical-physics-and-mathematical-physics/cosmic-strings-and-other-topological-defects?format=PB}
  {\emph {\bibinfo {title} {{Cosmic Strings and Other Topological Defects}}}}\
  (\bibinfo  {publisher} {Cambridge University Press},\ \bibinfo {year}
  {2000})\BibitemShut {NoStop}%
\bibitem [{\citenamefont {Manton}\ and\ \citenamefont
  {Sutcliffe}(2004)}]{Manton:2004tk}%
  \BibitemOpen
  \bibfield  {author} {\bibinfo {author} {\bibfnamefont {N.~S.}\ \bibnamefont
  {Manton}}\ and\ \bibinfo {author} {\bibfnamefont {P.}~\bibnamefont
  {Sutcliffe}},\ }\href {\doibase 10.1017/CBO9780511617034} {\emph {\bibinfo
  {title} {{Topological solitons}}}},\ Cambridge Monographs on Mathematical
  Physics\ (\bibinfo  {publisher} {Cambridge University Press},\ \bibinfo
  {year} {2004})\BibitemShut {NoStop}%
\bibitem [{\citenamefont {Weinberg}(2012)}]{Weinberg:2012pjx}%
  \BibitemOpen
  \bibfield  {author} {\bibinfo {author} {\bibfnamefont {E.~J.}\ \bibnamefont
  {Weinberg}},\ }\href {\doibase 10.1017/CBO9781139017787} {\emph {\bibinfo
  {title} {{Classical solutions in quantum field theory}}}},\ Cambridge
  Monographs on Mathematical Physics\ (\bibinfo  {publisher} {Cambridge
  University Press},\ \bibinfo {year} {2012})\BibitemShut {NoStop}%
\bibitem [{\citenamefont {Liebling}\ and\ \citenamefont
  {Palenzuela}(2012)}]{Liebling:2012fv}%
  \BibitemOpen
  \bibfield  {author} {\bibinfo {author} {\bibfnamefont {S.~L.}\ \bibnamefont
  {Liebling}}\ and\ \bibinfo {author} {\bibfnamefont {C.}~\bibnamefont
  {Palenzuela}},\ }\href {\doibase 10.12942/lrr-2012-6,
  10.1007/s41114-017-0007-y} {\bibfield  {journal} {\bibinfo  {journal} {Living
  Rev. Rel.}\ }\textbf {\bibinfo {volume} {15}},\ \bibinfo {pages} {6}
  (\bibinfo {year} {2012})},\ \bibinfo {note} {[Living Rev.
  Rel.20,no.1,5(2017)]},\ \Eprint {http://arxiv.org/abs/1202.5809}
  {arXiv:1202.5809 [gr-qc]} \BibitemShut {NoStop}%
\bibitem [{\citenamefont {Kivshar}\ and\ \citenamefont
  {Malomed}(1989)}]{RevModPhys.61.763}%
  \BibitemOpen
  \bibfield  {author} {\bibinfo {author} {\bibfnamefont {Y.~S.}\ \bibnamefont
  {Kivshar}}\ and\ \bibinfo {author} {\bibfnamefont {B.~A.}\ \bibnamefont
  {Malomed}},\ }\href {\doibase 10.1103/RevModPhys.61.763} {\bibfield
  {journal} {\bibinfo  {journal} {Rev. Mod. Phys.}\ }\textbf {\bibinfo {volume}
  {61}},\ \bibinfo {pages} {763} (\bibinfo {year} {1989})}\BibitemShut
  {NoStop}%
\bibitem [{\citenamefont {Amin}\ \emph {et~al.}(2012)\citenamefont {Amin},
  \citenamefont {Easther}, \citenamefont {Finkel}, \citenamefont {Flauger},\
  and\ \citenamefont {Hertzberg}}]{Amin:2011hj}%
  \BibitemOpen
  \bibfield  {author} {\bibinfo {author} {\bibfnamefont {M.~A.}\ \bibnamefont
  {Amin}}, \bibinfo {author} {\bibfnamefont {R.}~\bibnamefont {Easther}},
  \bibinfo {author} {\bibfnamefont {H.}~\bibnamefont {Finkel}}, \bibinfo
  {author} {\bibfnamefont {R.}~\bibnamefont {Flauger}}, \ and\ \bibinfo
  {author} {\bibfnamefont {M.~P.}\ \bibnamefont {Hertzberg}},\ }\href {\doibase
  10.1103/PhysRevLett.108.241302} {\bibfield  {journal} {\bibinfo  {journal}
  {Phys. Rev. Lett.}\ }\textbf {\bibinfo {volume} {108}},\ \bibinfo {pages}
  {241302} (\bibinfo {year} {2012})},\ \Eprint {http://arxiv.org/abs/1106.3335}
  {arXiv:1106.3335 [astro-ph.CO]} \BibitemShut {NoStop}%
\bibitem [{\citenamefont {Kolb}\ and\ \citenamefont
  {Tkachev}(1994)}]{Kolb:1993hw}%
  \BibitemOpen
  \bibfield  {author} {\bibinfo {author} {\bibfnamefont {E.~W.}\ \bibnamefont
  {Kolb}}\ and\ \bibinfo {author} {\bibfnamefont {I.~I.}\ \bibnamefont
  {Tkachev}},\ }\href {\doibase 10.1103/PhysRevD.49.5040} {\bibfield  {journal}
  {\bibinfo  {journal} {Phys. Rev.}\ }\textbf {\bibinfo {volume} {D49}},\
  \bibinfo {pages} {5040} (\bibinfo {year} {1994})},\ \Eprint
  {http://arxiv.org/abs/astro-ph/9311037} {arXiv:astro-ph/9311037 [astro-ph]}
  \BibitemShut {NoStop}%
\bibitem [{\citenamefont {Zhou}\ \emph {et~al.}(2013)\citenamefont {Zhou},
  \citenamefont {Copeland}, \citenamefont {Easther}, \citenamefont {Finkel},
  \citenamefont {Mou},\ and\ \citenamefont {Saffin}}]{Zhou:2013tsa}%
  \BibitemOpen
  \bibfield  {author} {\bibinfo {author} {\bibfnamefont {S.-Y.}\ \bibnamefont
  {Zhou}}, \bibinfo {author} {\bibfnamefont {E.~J.}\ \bibnamefont {Copeland}},
  \bibinfo {author} {\bibfnamefont {R.}~\bibnamefont {Easther}}, \bibinfo
  {author} {\bibfnamefont {H.}~\bibnamefont {Finkel}}, \bibinfo {author}
  {\bibfnamefont {Z.-G.}\ \bibnamefont {Mou}}, \ and\ \bibinfo {author}
  {\bibfnamefont {P.~M.}\ \bibnamefont {Saffin}},\ }\href {\doibase
  10.1007/JHEP10(2013)026} {\bibfield  {journal} {\bibinfo  {journal} {JHEP}\
  }\textbf {\bibinfo {volume} {10}},\ \bibinfo {pages} {026} (\bibinfo {year}
  {2013})},\ \Eprint {http://arxiv.org/abs/1304.6094} {arXiv:1304.6094
  [astro-ph.CO]} \BibitemShut {NoStop}%
\bibitem [{\citenamefont {Antusch}\ \emph {et~al.}(2018)\citenamefont
  {Antusch}, \citenamefont {Cefala}, \citenamefont {Krippendorf}, \citenamefont
  {Muia}, \citenamefont {Orani},\ and\ \citenamefont
  {Quevedo}}]{Antusch:2017flz}%
  \BibitemOpen
  \bibfield  {author} {\bibinfo {author} {\bibfnamefont {S.}~\bibnamefont
  {Antusch}}, \bibinfo {author} {\bibfnamefont {F.}~\bibnamefont {Cefala}},
  \bibinfo {author} {\bibfnamefont {S.}~\bibnamefont {Krippendorf}}, \bibinfo
  {author} {\bibfnamefont {F.}~\bibnamefont {Muia}}, \bibinfo {author}
  {\bibfnamefont {S.}~\bibnamefont {Orani}}, \ and\ \bibinfo {author}
  {\bibfnamefont {F.}~\bibnamefont {Quevedo}},\ }\href {\doibase
  10.1007/JHEP01(2018)083} {\bibfield  {journal} {\bibinfo  {journal} {JHEP}\
  }\textbf {\bibinfo {volume} {01}},\ \bibinfo {pages} {083} (\bibinfo {year}
  {2018})},\ \Eprint {http://arxiv.org/abs/1708.08922} {arXiv:1708.08922
  [hep-th]} \BibitemShut {NoStop}%
\bibitem [{\citenamefont {Helfer}\ \emph {et~al.}(2018)\citenamefont {Helfer},
  \citenamefont {Lim}, \citenamefont {Garcia},\ and\ \citenamefont
  {Amin}}]{Helfer:2018vtq}%
  \BibitemOpen
  \bibfield  {author} {\bibinfo {author} {\bibfnamefont {T.}~\bibnamefont
  {Helfer}}, \bibinfo {author} {\bibfnamefont {E.~A.}\ \bibnamefont {Lim}},
  \bibinfo {author} {\bibfnamefont {M.~A.~G.}\ \bibnamefont {Garcia}}, \ and\
  \bibinfo {author} {\bibfnamefont {M.~A.}\ \bibnamefont {Amin}},\ }\href@noop
  {} {\bibfield  {journal} {\bibinfo  {journal} {arXiv e-prints}\ } (\bibinfo
  {year} {2018})},\ \Eprint {http://arxiv.org/abs/1802.06733} {arXiv:1802.06733
  [gr-qc]} \BibitemShut {NoStop}%
\bibitem [{\citenamefont {Palenzuela}\ \emph {et~al.}(2017)\citenamefont
  {Palenzuela}, \citenamefont {Pani}, \citenamefont {Bezares}, \citenamefont
  {Cardoso}, \citenamefont {Lehner},\ and\ \citenamefont
  {Liebling}}]{Palenzuela:2017kcg}%
  \BibitemOpen
  \bibfield  {author} {\bibinfo {author} {\bibfnamefont {C.}~\bibnamefont
  {Palenzuela}}, \bibinfo {author} {\bibfnamefont {P.}~\bibnamefont {Pani}},
  \bibinfo {author} {\bibfnamefont {M.}~\bibnamefont {Bezares}}, \bibinfo
  {author} {\bibfnamefont {V.}~\bibnamefont {Cardoso}}, \bibinfo {author}
  {\bibfnamefont {L.}~\bibnamefont {Lehner}}, \ and\ \bibinfo {author}
  {\bibfnamefont {S.}~\bibnamefont {Liebling}},\ }\href {\doibase
  10.1103/PhysRevD.96.104058} {\bibfield  {journal} {\bibinfo  {journal} {Phys.
  Rev.}\ }\textbf {\bibinfo {volume} {D96}},\ \bibinfo {pages} {104058}
  (\bibinfo {year} {2017})},\ \Eprint {http://arxiv.org/abs/1710.09432}
  {arXiv:1710.09432 [gr-qc]} \BibitemShut {NoStop}%
\bibitem [{\citenamefont {Lozanov}\ and\ \citenamefont
  {Amin}(2019)}]{Lozanov:2019ylm}%
  \BibitemOpen
  \bibfield  {author} {\bibinfo {author} {\bibfnamefont {K.~D.}\ \bibnamefont
  {Lozanov}}\ and\ \bibinfo {author} {\bibfnamefont {M.~A.}\ \bibnamefont
  {Amin}},\ }\href {\doibase 10.1103/PhysRevD.99.123504} {\bibfield  {journal}
  {\bibinfo  {journal} {Phys. Rev.}\ }\textbf {\bibinfo {volume} {D99}},\
  \bibinfo {pages} {123504} (\bibinfo {year} {2019})},\ \Eprint
  {http://arxiv.org/abs/1902.06736} {arXiv:1902.06736 [astro-ph.CO]}
  \BibitemShut {NoStop}%
\bibitem [{\citenamefont {Khlopov}\ \emph {et~al.}(1985)\citenamefont
  {Khlopov}, \citenamefont {Malomed},\ and\ \citenamefont
  {Zeldovich}}]{Khlopov:1985jw}%
  \BibitemOpen
  \bibfield  {author} {\bibinfo {author} {\bibfnamefont {M.}~\bibnamefont
  {Khlopov}}, \bibinfo {author} {\bibfnamefont {B.~A.}\ \bibnamefont
  {Malomed}}, \ and\ \bibinfo {author} {\bibfnamefont {I.~B.}\ \bibnamefont
  {Zeldovich}},\ }\href@noop {} {\bibfield  {journal} {\bibinfo  {journal}
  {Mon. Not. Roy. Astron. Soc.}\ }\textbf {\bibinfo {volume} {215}},\ \bibinfo
  {pages} {575} (\bibinfo {year} {1985})}\BibitemShut {NoStop}%
\bibitem [{\citenamefont {Cotner}\ and\ \citenamefont
  {Kusenko}(2017)}]{Cotner:2016cvr}%
  \BibitemOpen
  \bibfield  {author} {\bibinfo {author} {\bibfnamefont {E.}~\bibnamefont
  {Cotner}}\ and\ \bibinfo {author} {\bibfnamefont {A.}~\bibnamefont
  {Kusenko}},\ }\href {\doibase 10.1103/PhysRevLett.119.031103} {\bibfield
  {journal} {\bibinfo  {journal} {Phys. Rev. Lett.}\ }\textbf {\bibinfo
  {volume} {119}},\ \bibinfo {pages} {031103} (\bibinfo {year} {2017})},\
  \Eprint {http://arxiv.org/abs/1612.02529} {arXiv:1612.02529 [astro-ph.CO]}
  \BibitemShut {NoStop}%
\bibitem [{\citenamefont {Cotner}\ \emph {et~al.}(2018)\citenamefont {Cotner},
  \citenamefont {Kusenko},\ and\ \citenamefont {Takhistov}}]{Cotner:2018vug}%
  \BibitemOpen
  \bibfield  {author} {\bibinfo {author} {\bibfnamefont {E.}~\bibnamefont
  {Cotner}}, \bibinfo {author} {\bibfnamefont {A.}~\bibnamefont {Kusenko}}, \
  and\ \bibinfo {author} {\bibfnamefont {V.}~\bibnamefont {Takhistov}},\ }\href
  {\doibase 10.1103/PhysRevD.98.083513} {\bibfield  {journal} {\bibinfo
  {journal} {Phys. Rev.}\ }\textbf {\bibinfo {volume} {D98}},\ \bibinfo {pages}
  {083513} (\bibinfo {year} {2018})},\ \Eprint
  {http://arxiv.org/abs/1801.03321} {arXiv:1801.03321 [astro-ph.CO]}
  \BibitemShut {NoStop}%
\bibitem [{\citenamefont {Enqvist}\ and\ \citenamefont
  {McDonald}(1998)}]{Enqvist:1997si}%
  \BibitemOpen
  \bibfield  {author} {\bibinfo {author} {\bibfnamefont {K.}~\bibnamefont
  {Enqvist}}\ and\ \bibinfo {author} {\bibfnamefont {J.}~\bibnamefont
  {McDonald}},\ }\href {\doibase 10.1016/S0370-2693(98)00271-8} {\bibfield
  {journal} {\bibinfo  {journal} {Phys. Lett.}\ }\textbf {\bibinfo {volume}
  {B425}},\ \bibinfo {pages} {309} (\bibinfo {year} {1998})},\ \Eprint
  {http://arxiv.org/abs/hep-ph/9711514} {arXiv:hep-ph/9711514 [hep-ph]}
  \BibitemShut {NoStop}%
\bibitem [{\citenamefont {Lozanov}\ and\ \citenamefont {Amin}(2014)}]{LozAmin}%
  \BibitemOpen
  \bibfield  {author} {\bibinfo {author} {\bibfnamefont {K.~D.}\ \bibnamefont
  {Lozanov}}\ and\ \bibinfo {author} {\bibfnamefont {M.~A.}\ \bibnamefont
  {Amin}},\ }\href {\doibase 10.1103/PhysRevD.90.083528} {\bibfield  {journal}
  {\bibinfo  {journal} {Phys.Rev.}\ }\textbf {\bibinfo {volume} {D90}},\
  \bibinfo {pages} {083528} (\bibinfo {year} {2014})},\ \Eprint
  {http://arxiv.org/abs/1408.1811} {arXiv:1408.1811 [hep-ph]} \BibitemShut
  {NoStop}%
\bibitem [{\citenamefont {Adshead}\ \emph {et~al.}(2015)\citenamefont
  {Adshead}, \citenamefont {Giblin}, \citenamefont {Scully},\ and\
  \citenamefont {Sfakianakis}}]{Adshead:2015pva}%
  \BibitemOpen
  \bibfield  {author} {\bibinfo {author} {\bibfnamefont {P.}~\bibnamefont
  {Adshead}}, \bibinfo {author} {\bibfnamefont {J.~T.}\ \bibnamefont {Giblin}},
  \bibinfo {author} {\bibfnamefont {T.~R.}\ \bibnamefont {Scully}}, \ and\
  \bibinfo {author} {\bibfnamefont {E.~I.}\ \bibnamefont {Sfakianakis}},\
  }\href {\doibase 10.1088/1475-7516/2015/12/034} {\bibfield  {journal}
  {\bibinfo  {journal} {JCAP}\ }\textbf {\bibinfo {volume} {1512}},\ \bibinfo
  {pages} {034} (\bibinfo {year} {2015})},\ \Eprint
  {http://arxiv.org/abs/1502.06506} {arXiv:1502.06506 [astro-ph.CO]}
  \BibitemShut {NoStop}%
\bibitem [{\citenamefont {Lozanov}\ and\ \citenamefont
  {Amin}(2017)}]{Lozanov:2016hid}%
  \BibitemOpen
  \bibfield  {author} {\bibinfo {author} {\bibfnamefont {K.~D.}\ \bibnamefont
  {Lozanov}}\ and\ \bibinfo {author} {\bibfnamefont {M.~A.}\ \bibnamefont
  {Amin}},\ }\href {\doibase 10.1103/PhysRevLett.119.061301} {\bibfield
  {journal} {\bibinfo  {journal} {Phys. Rev. Lett.}\ }\textbf {\bibinfo
  {volume} {119}},\ \bibinfo {pages} {061301} (\bibinfo {year} {2017})},\
  \Eprint {http://arxiv.org/abs/1608.01213} {arXiv:1608.01213 [astro-ph.CO]}
  \BibitemShut {NoStop}%
\bibitem [{\citenamefont {Lozanov}\ and\ \citenamefont
  {Amin}(2018)}]{Lozanov:2017hjm}%
  \BibitemOpen
  \bibfield  {author} {\bibinfo {author} {\bibfnamefont {K.~D.}\ \bibnamefont
  {Lozanov}}\ and\ \bibinfo {author} {\bibfnamefont {M.~A.}\ \bibnamefont
  {Amin}},\ }\href {\doibase 10.1103/PhysRevD.97.023533} {\bibfield  {journal}
  {\bibinfo  {journal} {Phys. Rev.}\ }\textbf {\bibinfo {volume} {D97}},\
  \bibinfo {pages} {023533} (\bibinfo {year} {2018})},\ \Eprint
  {http://arxiv.org/abs/1710.06851} {arXiv:1710.06851 [astro-ph.CO]}
  \BibitemShut {NoStop}%
\bibitem [{\citenamefont {Hu}\ \emph {et~al.}(2000)\citenamefont {Hu},
  \citenamefont {Barkana},\ and\ \citenamefont {Gruzinov}}]{Hu:2000ke}%
  \BibitemOpen
  \bibfield  {author} {\bibinfo {author} {\bibfnamefont {W.}~\bibnamefont
  {Hu}}, \bibinfo {author} {\bibfnamefont {R.}~\bibnamefont {Barkana}}, \ and\
  \bibinfo {author} {\bibfnamefont {A.}~\bibnamefont {Gruzinov}},\ }\href
  {\doibase 10.1103/PhysRevLett.85.1158} {\bibfield  {journal} {\bibinfo
  {journal} {Phys. Rev. Lett.}\ }\textbf {\bibinfo {volume} {85}},\ \bibinfo
  {pages} {1158} (\bibinfo {year} {2000})},\ \Eprint
  {http://arxiv.org/abs/astro-ph/0003365} {arXiv:astro-ph/0003365 [astro-ph]}
  \BibitemShut {NoStop}%
\bibitem [{\citenamefont {Alcubierre}\ \emph {et~al.}(2002)\citenamefont
  {Alcubierre}, \citenamefont {Guzman}, \citenamefont {Matos}, \citenamefont
  {Nunez}, \citenamefont {Urena-Lopez},\ and\ \citenamefont
  {Wiederhold}}]{Alcubierre:2001ea}%
  \BibitemOpen
  \bibfield  {author} {\bibinfo {author} {\bibfnamefont {M.}~\bibnamefont
  {Alcubierre}}, \bibinfo {author} {\bibfnamefont {F.~S.}\ \bibnamefont
  {Guzman}}, \bibinfo {author} {\bibfnamefont {T.}~\bibnamefont {Matos}},
  \bibinfo {author} {\bibfnamefont {D.}~\bibnamefont {Nunez}}, \bibinfo
  {author} {\bibfnamefont {L.~A.}\ \bibnamefont {Urena-Lopez}}, \ and\ \bibinfo
  {author} {\bibfnamefont {P.}~\bibnamefont {Wiederhold}},\ }\href {\doibase
  10.1088/0264-9381/19/19/314} {\bibfield  {journal} {\bibinfo  {journal}
  {Class. Quant. Grav.}\ }\textbf {\bibinfo {volume} {19}},\ \bibinfo {pages}
  {5017} (\bibinfo {year} {2002})},\ \Eprint
  {http://arxiv.org/abs/gr-qc/0110102} {arXiv:gr-qc/0110102 [gr-qc]}
  \BibitemShut {NoStop}%
\bibitem [{\citenamefont {Marsh}\ and\ \citenamefont
  {Pop}(2015)}]{Marsh:2015wka}%
  \BibitemOpen
  \bibfield  {author} {\bibinfo {author} {\bibfnamefont {D.~J.~E.}\
  \bibnamefont {Marsh}}\ and\ \bibinfo {author} {\bibfnamefont {A.-R.}\
  \bibnamefont {Pop}},\ }\href {\doibase 10.1093/mnras/stv1050} {\bibfield
  {journal} {\bibinfo  {journal} {Mon. Not. Roy. Astron. Soc.}\ }\textbf
  {\bibinfo {volume} {451}},\ \bibinfo {pages} {2479} (\bibinfo {year}
  {2015})},\ \Eprint {http://arxiv.org/abs/1502.03456} {arXiv:1502.03456
  [astro-ph.CO]} \BibitemShut {NoStop}%
\bibitem [{\citenamefont {Schive}\ \emph {et~al.}(2014)\citenamefont {Schive},
  \citenamefont {Chiueh},\ and\ \citenamefont {Broadhurst}}]{Schive:2014dra}%
  \BibitemOpen
  \bibfield  {author} {\bibinfo {author} {\bibfnamefont {H.-Y.}\ \bibnamefont
  {Schive}}, \bibinfo {author} {\bibfnamefont {T.}~\bibnamefont {Chiueh}}, \
  and\ \bibinfo {author} {\bibfnamefont {T.}~\bibnamefont {Broadhurst}},\
  }\href {\doibase 10.1038/nphys2996} {\bibfield  {journal} {\bibinfo
  {journal} {Nature Phys.}\ }\textbf {\bibinfo {volume} {10}},\ \bibinfo
  {pages} {496} (\bibinfo {year} {2014})},\ \Eprint
  {http://arxiv.org/abs/1406.6586} {arXiv:1406.6586 [astro-ph.GA]} \BibitemShut
  {NoStop}%
\bibitem [{\citenamefont {Hui}\ \emph {et~al.}(2017)\citenamefont {Hui},
  \citenamefont {Ostriker}, \citenamefont {Tremaine},\ and\ \citenamefont
  {Witten}}]{Hui:2016ltb}%
  \BibitemOpen
  \bibfield  {author} {\bibinfo {author} {\bibfnamefont {L.}~\bibnamefont
  {Hui}}, \bibinfo {author} {\bibfnamefont {J.~P.}\ \bibnamefont {Ostriker}},
  \bibinfo {author} {\bibfnamefont {S.}~\bibnamefont {Tremaine}}, \ and\
  \bibinfo {author} {\bibfnamefont {E.}~\bibnamefont {Witten}},\ }\href
  {\doibase 10.1103/PhysRevD.95.043541} {\bibfield  {journal} {\bibinfo
  {journal} {Phys. Rev.}\ }\textbf {\bibinfo {volume} {D95}},\ \bibinfo {pages}
  {043541} (\bibinfo {year} {2017})},\ \Eprint
  {http://arxiv.org/abs/1610.08297} {arXiv:1610.08297 [astro-ph.CO]}
  \BibitemShut {NoStop}%
\bibitem [{\citenamefont {Kaup}(1968)}]{Kaup:1968}%
  \BibitemOpen
  \bibfield  {author} {\bibinfo {author} {\bibfnamefont {D.~J.}\ \bibnamefont
  {Kaup}},\ }\href {\doibase 10.1103/PhysRev.172.1331} {\bibfield  {journal}
  {\bibinfo  {journal} {Phys. Rev.}\ }\textbf {\bibinfo {volume} {172}},\
  \bibinfo {pages} {1331} (\bibinfo {year} {1968})}\BibitemShut {NoStop}%
\bibitem [{\citenamefont {Coleman}(1985)}]{Coleman:1985ki}%
  \BibitemOpen
  \bibfield  {author} {\bibinfo {author} {\bibfnamefont {S.~R.}\ \bibnamefont
  {Coleman}},\ }\href {\doibase 10.1016/0550-3213(85)90286-X,
  10.1016/0550-3213(86)90520-1} {\bibfield  {journal} {\bibinfo  {journal}
  {Nucl. Phys.}\ }\textbf {\bibinfo {volume} {B262}},\ \bibinfo {pages} {263}
  (\bibinfo {year} {1985})},\ \bibinfo {note} {[Erratum: Nucl.
  Phys.B269,744(1986)]}\BibitemShut {NoStop}%
\bibitem [{\citenamefont {Bogolyubsky}\ and\ \citenamefont
  {Makhankov}(1976)}]{Bogolyubsky:1976nx}%
  \BibitemOpen
  \bibfield  {author} {\bibinfo {author} {\bibfnamefont {I.~L.}\ \bibnamefont
  {Bogolyubsky}}\ and\ \bibinfo {author} {\bibfnamefont {V.~G.}\ \bibnamefont
  {Makhankov}},\ }\href@noop {} {\bibfield  {journal} {\bibinfo  {journal}
  {JETP Lett.}\ }\textbf {\bibinfo {volume} {24}},\ \bibinfo {pages} {12}
  (\bibinfo {year} {1976})}\BibitemShut {NoStop}%
\bibitem [{\citenamefont {Copeland}\ \emph {et~al.}(1995)\citenamefont
  {Copeland}, \citenamefont {Gleiser},\ and\ \citenamefont
  {Muller}}]{Copeland:1995fq}%
  \BibitemOpen
  \bibfield  {author} {\bibinfo {author} {\bibfnamefont {E.~J.}\ \bibnamefont
  {Copeland}}, \bibinfo {author} {\bibfnamefont {M.}~\bibnamefont {Gleiser}}, \
  and\ \bibinfo {author} {\bibfnamefont {H.~R.}\ \bibnamefont {Muller}},\
  }\href {\doibase 10.1103/PhysRevD.52.1920} {\bibfield  {journal} {\bibinfo
  {journal} {Phys. Rev.}\ }\textbf {\bibinfo {volume} {D52}},\ \bibinfo {pages}
  {1920} (\bibinfo {year} {1995})},\ \Eprint
  {http://arxiv.org/abs/hep-ph/9503217} {arXiv:hep-ph/9503217 [hep-ph]}
  \BibitemShut {NoStop}%
\bibitem [{\citenamefont {Amin}\ and\ \citenamefont
  {Shirokoff}(2010)}]{Amin:2010jq}%
  \BibitemOpen
  \bibfield  {author} {\bibinfo {author} {\bibfnamefont {M.~A.}\ \bibnamefont
  {Amin}}\ and\ \bibinfo {author} {\bibfnamefont {D.}~\bibnamefont
  {Shirokoff}},\ }\href {\doibase 10.1103/PhysRevD.81.085045} {\bibfield
  {journal} {\bibinfo  {journal} {Phys. Rev.}\ }\textbf {\bibinfo {volume}
  {D81}},\ \bibinfo {pages} {085045} (\bibinfo {year} {2010})},\ \Eprint
  {http://arxiv.org/abs/1002.3380} {arXiv:1002.3380 [astro-ph.CO]} \BibitemShut
  {NoStop}%
\bibitem [{\citenamefont {Kasuya}\ \emph {et~al.}(2003)\citenamefont {Kasuya},
  \citenamefont {Kawasaki},\ and\ \citenamefont {Takahashi}}]{Kasuya:2002zs}%
  \BibitemOpen
  \bibfield  {author} {\bibinfo {author} {\bibfnamefont {S.}~\bibnamefont
  {Kasuya}}, \bibinfo {author} {\bibfnamefont {M.}~\bibnamefont {Kawasaki}}, \
  and\ \bibinfo {author} {\bibfnamefont {F.}~\bibnamefont {Takahashi}},\ }\href
  {\doibase 10.1016/S0370-2693(03)00344-7} {\bibfield  {journal} {\bibinfo
  {journal} {Phys. Lett.}\ }\textbf {\bibinfo {volume} {B559}},\ \bibinfo
  {pages} {99} (\bibinfo {year} {2003})},\ \Eprint
  {http://arxiv.org/abs/hep-ph/0209358} {arXiv:hep-ph/0209358 [hep-ph]}
  \BibitemShut {NoStop}%
\bibitem [{\citenamefont {Amin}(2013)}]{Amin:2013ika}%
  \BibitemOpen
  \bibfield  {author} {\bibinfo {author} {\bibfnamefont {M.~A.}\ \bibnamefont
  {Amin}},\ }\href {\doibase 10.1103/PhysRevD.87.123505} {\bibfield  {journal}
  {\bibinfo  {journal} {Phys. Rev.}\ }\textbf {\bibinfo {volume} {D87}},\
  \bibinfo {pages} {123505} (\bibinfo {year} {2013})},\ \Eprint
  {http://arxiv.org/abs/1303.1102} {arXiv:1303.1102 [astro-ph.CO]} \BibitemShut
  {NoStop}%
\bibitem [{\citenamefont {{Chavanis}}(2011)}]{2011PhRvD..84d3531C}%
  \BibitemOpen
  \bibfield  {author} {\bibinfo {author} {\bibfnamefont {P.-H.}\ \bibnamefont
  {{Chavanis}}},\ }\href {\doibase 10.1103/PhysRevD.84.043531} {\bibfield
  {journal} {\bibinfo  {journal} {\prd}\ }\textbf {\bibinfo {volume} {84}},\
  \bibinfo {eid} {043531} (\bibinfo {year} {2011})},\ \Eprint
  {http://arxiv.org/abs/1103.2050} {arXiv:1103.2050} \BibitemShut {NoStop}%
\bibitem [{\citenamefont {{Chavanis}}\ and\ \citenamefont
  {{Delfini}}(2011)}]{2011PhRvD..84d3532C}%
  \BibitemOpen
  \bibfield  {author} {\bibinfo {author} {\bibfnamefont {P.-H.}\ \bibnamefont
  {{Chavanis}}}\ and\ \bibinfo {author} {\bibfnamefont {L.}~\bibnamefont
  {{Delfini}}},\ }\href {\doibase 10.1103/PhysRevD.84.043532} {\bibfield
  {journal} {\bibinfo  {journal} {\prd}\ }\textbf {\bibinfo {volume} {84}},\
  \bibinfo {eid} {043532} (\bibinfo {year} {2011})},\ \Eprint
  {http://arxiv.org/abs/1103.2054} {arXiv:1103.2054 [astro-ph.CO]} \BibitemShut
  {NoStop}%
\bibitem [{\citenamefont {{Chavanis}}(2012)}]{2012A&A...537A.127C}%
  \BibitemOpen
  \bibfield  {author} {\bibinfo {author} {\bibfnamefont {P.~H.}\ \bibnamefont
  {{Chavanis}}},\ }\href {\doibase 10.1051/0004-6361/201116905} {\bibfield
  {journal} {\bibinfo  {journal} {\aap}\ }\textbf {\bibinfo {volume} {537}},\
  \bibinfo {eid} {A127} (\bibinfo {year} {2012})},\ \Eprint
  {http://arxiv.org/abs/1103.2698} {arXiv:1103.2698} \BibitemShut {NoStop}%
\bibitem [{\citenamefont {{Chavanis}}(2018)}]{2018PhRvD..98b3009C}%
  \BibitemOpen
  \bibfield  {author} {\bibinfo {author} {\bibfnamefont {P.-H.}\ \bibnamefont
  {{Chavanis}}},\ }\href {\doibase 10.1103/PhysRevD.98.023009} {\bibfield
  {journal} {\bibinfo  {journal} {\prd}\ }\textbf {\bibinfo {volume} {98}},\
  \bibinfo {eid} {023009} (\bibinfo {year} {2018})},\ \Eprint
  {http://arxiv.org/abs/1710.06268} {arXiv:1710.06268 [gr-qc]} \BibitemShut
  {NoStop}%
\bibitem [{\citenamefont {Kusenko}\ and\ \citenamefont
  {Shaposhnikov}(1998)}]{Kusenko:1997si}%
  \BibitemOpen
  \bibfield  {author} {\bibinfo {author} {\bibfnamefont {A.}~\bibnamefont
  {Kusenko}}\ and\ \bibinfo {author} {\bibfnamefont {M.~E.}\ \bibnamefont
  {Shaposhnikov}},\ }\href {\doibase 10.1016/S0370-2693(97)01375-0} {\bibfield
  {journal} {\bibinfo  {journal} {Phys. Lett.}\ }\textbf {\bibinfo {volume}
  {B418}},\ \bibinfo {pages} {46} (\bibinfo {year} {1998})},\ \Eprint
  {http://arxiv.org/abs/hep-ph/9709492} {arXiv:hep-ph/9709492 [hep-ph]}
  \BibitemShut {NoStop}%
\bibitem [{\citenamefont {Farhi}\ \emph {et~al.}(2008)\citenamefont {Farhi},
  \citenamefont {Graham}, \citenamefont {Guth}, \citenamefont {Iqbal},
  \citenamefont {Rosales},\ and\ \citenamefont {Stamatopoulos}}]{Farhi:2007wj}%
  \BibitemOpen
  \bibfield  {author} {\bibinfo {author} {\bibfnamefont {E.}~\bibnamefont
  {Farhi}}, \bibinfo {author} {\bibfnamefont {N.}~\bibnamefont {Graham}},
  \bibinfo {author} {\bibfnamefont {A.~H.}\ \bibnamefont {Guth}}, \bibinfo
  {author} {\bibfnamefont {N.}~\bibnamefont {Iqbal}}, \bibinfo {author}
  {\bibfnamefont {R.~R.}\ \bibnamefont {Rosales}}, \ and\ \bibinfo {author}
  {\bibfnamefont {N.}~\bibnamefont {Stamatopoulos}},\ }\href {\doibase
  10.1103/PhysRevD.77.085019} {\bibfield  {journal} {\bibinfo  {journal} {Phys.
  Rev.}\ }\textbf {\bibinfo {volume} {D77}},\ \bibinfo {pages} {085019}
  (\bibinfo {year} {2008})},\ \Eprint {http://arxiv.org/abs/0712.3034}
  {arXiv:0712.3034 [hep-th]} \BibitemShut {NoStop}%
\bibitem [{\citenamefont {Gleiser}\ \emph {et~al.}(2011)\citenamefont
  {Gleiser}, \citenamefont {Graham},\ and\ \citenamefont
  {Stamatopoulos}}]{Gleiser:2011xj}%
  \BibitemOpen
  \bibfield  {author} {\bibinfo {author} {\bibfnamefont {M.}~\bibnamefont
  {Gleiser}}, \bibinfo {author} {\bibfnamefont {N.}~\bibnamefont {Graham}}, \
  and\ \bibinfo {author} {\bibfnamefont {N.}~\bibnamefont {Stamatopoulos}},\
  }\href {\doibase 10.1103/PhysRevD.83.096010} {\bibfield  {journal} {\bibinfo
  {journal} {Phys. Rev.}\ }\textbf {\bibinfo {volume} {D83}},\ \bibinfo {pages}
  {096010} (\bibinfo {year} {2011})},\ \Eprint {http://arxiv.org/abs/1103.1911}
  {arXiv:1103.1911 [hep-th]} \BibitemShut {NoStop}%
\bibitem [{\citenamefont {{Mocz}}\ \emph {et~al.}(2017)\citenamefont {{Mocz}},
  \citenamefont {{Vogelsberger}}, \citenamefont {{Robles}}, \citenamefont
  {{Zavala}}, \citenamefont {{Boylan-Kolchin}}, \citenamefont {{Fialkov}},\
  and\ \citenamefont {{Hernquist}}}]{2017MNRAS.471.4559M}%
  \BibitemOpen
  \bibfield  {author} {\bibinfo {author} {\bibfnamefont {P.}~\bibnamefont
  {{Mocz}}}, \bibinfo {author} {\bibfnamefont {M.}~\bibnamefont
  {{Vogelsberger}}}, \bibinfo {author} {\bibfnamefont {V.~H.}\ \bibnamefont
  {{Robles}}}, \bibinfo {author} {\bibfnamefont {J.}~\bibnamefont {{Zavala}}},
  \bibinfo {author} {\bibfnamefont {M.}~\bibnamefont {{Boylan-Kolchin}}},
  \bibinfo {author} {\bibfnamefont {A.}~\bibnamefont {{Fialkov}}}, \ and\
  \bibinfo {author} {\bibfnamefont {L.}~\bibnamefont {{Hernquist}}},\ }\href
  {\doibase 10.1093/mnras/stx1887} {\bibfield  {journal} {\bibinfo  {journal}
  {\mnras}\ }\textbf {\bibinfo {volume} {471}},\ \bibinfo {pages} {4559}
  (\bibinfo {year} {2017})},\ \Eprint {http://arxiv.org/abs/1705.05845}
  {arXiv:1705.05845} \BibitemShut {NoStop}%
\bibitem [{\citenamefont {{Schwabe}}\ \emph {et~al.}(2016)\citenamefont
  {{Schwabe}}, \citenamefont {{Niemeyer}},\ and\ \citenamefont
  {{Engels}}}]{2016PhRvD..94d3513S}%
  \BibitemOpen
  \bibfield  {author} {\bibinfo {author} {\bibfnamefont {B.}~\bibnamefont
  {{Schwabe}}}, \bibinfo {author} {\bibfnamefont {J.~C.}\ \bibnamefont
  {{Niemeyer}}}, \ and\ \bibinfo {author} {\bibfnamefont {J.~F.}\ \bibnamefont
  {{Engels}}},\ }\href {\doibase 10.1103/PhysRevD.94.043513} {\bibfield
  {journal} {\bibinfo  {journal} {\prd}\ }\textbf {\bibinfo {volume} {94}},\
  \bibinfo {eid} {043513} (\bibinfo {year} {2016})},\ \Eprint
  {http://arxiv.org/abs/1606.05151} {arXiv:1606.05151} \BibitemShut {NoStop}%
\bibitem [{\citenamefont {Palenzuela}\ \emph {et~al.}(2007)\citenamefont
  {Palenzuela}, \citenamefont {Olabarrieta}, \citenamefont {Lehner},\ and\
  \citenamefont {Liebling}}]{Palenzuela:2006wp}%
  \BibitemOpen
  \bibfield  {author} {\bibinfo {author} {\bibfnamefont {C.}~\bibnamefont
  {Palenzuela}}, \bibinfo {author} {\bibfnamefont {I.}~\bibnamefont
  {Olabarrieta}}, \bibinfo {author} {\bibfnamefont {L.}~\bibnamefont {Lehner}},
  \ and\ \bibinfo {author} {\bibfnamefont {S.~L.}\ \bibnamefont {Liebling}},\
  }\href {\doibase 10.1103/PhysRevD.75.064005} {\bibfield  {journal} {\bibinfo
  {journal} {Phys. Rev.}\ }\textbf {\bibinfo {volume} {D75}},\ \bibinfo {pages}
  {064005} (\bibinfo {year} {2007})},\ \Eprint
  {http://arxiv.org/abs/gr-qc/0612067} {arXiv:gr-qc/0612067 [gr-qc]}
  \BibitemShut {NoStop}%
\bibitem [{\citenamefont {Amin}\ \emph {et~al.}(2014)\citenamefont {Amin},
  \citenamefont {Banik}, \citenamefont {Negreanu},\ and\ \citenamefont
  {Yang}}]{Amin:2014fua}%
  \BibitemOpen
  \bibfield  {author} {\bibinfo {author} {\bibfnamefont {M.~A.}\ \bibnamefont
  {Amin}}, \bibinfo {author} {\bibfnamefont {I.}~\bibnamefont {Banik}},
  \bibinfo {author} {\bibfnamefont {C.}~\bibnamefont {Negreanu}}, \ and\
  \bibinfo {author} {\bibfnamefont {I.-S.}\ \bibnamefont {Yang}},\ }\href
  {\doibase 10.1103/PhysRevD.90.085024} {\bibfield  {journal} {\bibinfo
  {journal} {Phys. Rev.}\ }\textbf {\bibinfo {volume} {D90}},\ \bibinfo {pages}
  {085024} (\bibinfo {year} {2014})},\ \Eprint {http://arxiv.org/abs/1410.1822}
  {arXiv:1410.1822 [hep-th]} \BibitemShut {NoStop}%
\bibitem [{\citenamefont {Hindmarsh}\ and\ \citenamefont
  {Salmi}(2008)}]{Hindmarsh:2007jb}%
  \BibitemOpen
  \bibfield  {author} {\bibinfo {author} {\bibfnamefont {M.}~\bibnamefont
  {Hindmarsh}}\ and\ \bibinfo {author} {\bibfnamefont {P.}~\bibnamefont
  {Salmi}},\ }\href {\doibase 10.1103/PhysRevD.77.105025} {\bibfield  {journal}
  {\bibinfo  {journal} {Phys. Rev.}\ }\textbf {\bibinfo {volume} {D77}},\
  \bibinfo {pages} {105025} (\bibinfo {year} {2008})},\ \Eprint
  {http://arxiv.org/abs/0712.0614} {arXiv:0712.0614 [hep-th]} \BibitemShut
  {NoStop}%
\bibitem [{\citenamefont {Schwabe}\ \emph {et~al.}(2016)\citenamefont
  {Schwabe}, \citenamefont {Niemeyer},\ and\ \citenamefont
  {Engels}}]{Schwabe:2016rze}%
  \BibitemOpen
  \bibfield  {author} {\bibinfo {author} {\bibfnamefont {B.}~\bibnamefont
  {Schwabe}}, \bibinfo {author} {\bibfnamefont {J.~C.}\ \bibnamefont
  {Niemeyer}}, \ and\ \bibinfo {author} {\bibfnamefont {J.~F.}\ \bibnamefont
  {Engels}},\ }\href {\doibase 10.1103/PhysRevD.94.043513} {\bibfield
  {journal} {\bibinfo  {journal} {Phys. Rev.}\ }\textbf {\bibinfo {volume}
  {D94}},\ \bibinfo {pages} {043513} (\bibinfo {year} {2016})},\ \Eprint
  {http://arxiv.org/abs/1606.05151} {arXiv:1606.05151 [astro-ph.CO]}
  \BibitemShut {NoStop}%
\bibitem [{\citenamefont {Arvanitaki}\ \emph {et~al.}(2010)\citenamefont
  {Arvanitaki}, \citenamefont {Dimopoulos}, \citenamefont {Dubovsky},
  \citenamefont {Kaloper},\ and\ \citenamefont
  {March-Russell}}]{Arvanitaki:2009fg}%
  \BibitemOpen
  \bibfield  {author} {\bibinfo {author} {\bibfnamefont {A.}~\bibnamefont
  {Arvanitaki}}, \bibinfo {author} {\bibfnamefont {S.}~\bibnamefont
  {Dimopoulos}}, \bibinfo {author} {\bibfnamefont {S.}~\bibnamefont
  {Dubovsky}}, \bibinfo {author} {\bibfnamefont {N.}~\bibnamefont {Kaloper}}, \
  and\ \bibinfo {author} {\bibfnamefont {J.}~\bibnamefont {March-Russell}},\
  }\href {\doibase 10.1103/PhysRevD.81.123530} {\bibfield  {journal} {\bibinfo
  {journal} {Phys. Rev.}\ }\textbf {\bibinfo {volume} {D81}},\ \bibinfo {pages}
  {123530} (\bibinfo {year} {2010})},\ \Eprint {http://arxiv.org/abs/0905.4720}
  {arXiv:0905.4720 [hep-th]} \BibitemShut {NoStop}%
\bibitem [{\citenamefont {Hlozek}\ \emph {et~al.}(2015)\citenamefont {Hlozek},
  \citenamefont {Grin}, \citenamefont {Marsh},\ and\ \citenamefont
  {Ferreira}}]{Hlozek:2014lca}%
  \BibitemOpen
  \bibfield  {author} {\bibinfo {author} {\bibfnamefont {R.}~\bibnamefont
  {Hlozek}}, \bibinfo {author} {\bibfnamefont {D.}~\bibnamefont {Grin}},
  \bibinfo {author} {\bibfnamefont {D.~J.~E.}\ \bibnamefont {Marsh}}, \ and\
  \bibinfo {author} {\bibfnamefont {P.~G.}\ \bibnamefont {Ferreira}},\ }\href
  {\doibase 10.1103/PhysRevD.91.103512} {\bibfield  {journal} {\bibinfo
  {journal} {Phys. Rev.}\ }\textbf {\bibinfo {volume} {D91}},\ \bibinfo {pages}
  {103512} (\bibinfo {year} {2015})},\ \Eprint {http://arxiv.org/abs/1410.2896}
  {arXiv:1410.2896 [astro-ph.CO]} \BibitemShut {NoStop}%
\bibitem [{\citenamefont {Namjoo}\ \emph {et~al.}(2018)\citenamefont {Namjoo},
  \citenamefont {Guth},\ and\ \citenamefont {Kaiser}}]{Namjoo:2017nia}%
  \BibitemOpen
  \bibfield  {author} {\bibinfo {author} {\bibfnamefont {M.~H.}\ \bibnamefont
  {Namjoo}}, \bibinfo {author} {\bibfnamefont {A.~H.}\ \bibnamefont {Guth}}, \
  and\ \bibinfo {author} {\bibfnamefont {D.~I.}\ \bibnamefont {Kaiser}},\
  }\href {\doibase 10.1103/PhysRevD.98.016011} {\bibfield  {journal} {\bibinfo
  {journal} {Phys. Rev.}\ }\textbf {\bibinfo {volume} {D98}},\ \bibinfo {pages}
  {016011} (\bibinfo {year} {2018})},\ \Eprint
  {http://arxiv.org/abs/1712.00445} {arXiv:1712.00445 [hep-ph]} \BibitemShut
  {NoStop}%
\bibitem [{\citenamefont {Eby}\ \emph {et~al.}(2018)\citenamefont {Eby},
  \citenamefont {Mukaida}, \citenamefont {Takimoto}, \citenamefont
  {Wijewardhana},\ and\ \citenamefont {Yamada}}]{Eby:2018ufi}%
  \BibitemOpen
  \bibfield  {author} {\bibinfo {author} {\bibfnamefont {J.}~\bibnamefont
  {Eby}}, \bibinfo {author} {\bibfnamefont {K.}~\bibnamefont {Mukaida}},
  \bibinfo {author} {\bibfnamefont {M.}~\bibnamefont {Takimoto}}, \bibinfo
  {author} {\bibfnamefont {L.~C.~R.}\ \bibnamefont {Wijewardhana}}, \ and\
  \bibinfo {author} {\bibfnamefont {M.}~\bibnamefont {Yamada}},\ }\href@noop {}
  {\bibfield  {journal} {\bibinfo  {journal} {arXiv e-prints}\ } (\bibinfo
  {year} {2018})},\ \Eprint {http://arxiv.org/abs/1807.09795} {arXiv:1807.09795
  [hep-ph]} \BibitemShut {NoStop}%
\bibitem [{\citenamefont {Mocz}\ \emph {et~al.}(2017)\citenamefont {Mocz},
  \citenamefont {Vogelsberger}, \citenamefont {Robles}, \citenamefont {Zavala},
  \citenamefont {Boylan-Kolchin}, \citenamefont {Fialkov},\ and\ \citenamefont
  {Hernquist}}]{Mocz:2017wlg}%
  \BibitemOpen
  \bibfield  {author} {\bibinfo {author} {\bibfnamefont {P.}~\bibnamefont
  {Mocz}}, \bibinfo {author} {\bibfnamefont {M.}~\bibnamefont {Vogelsberger}},
  \bibinfo {author} {\bibfnamefont {V.~H.}\ \bibnamefont {Robles}}, \bibinfo
  {author} {\bibfnamefont {J.}~\bibnamefont {Zavala}}, \bibinfo {author}
  {\bibfnamefont {M.}~\bibnamefont {Boylan-Kolchin}}, \bibinfo {author}
  {\bibfnamefont {A.}~\bibnamefont {Fialkov}}, \ and\ \bibinfo {author}
  {\bibfnamefont {L.}~\bibnamefont {Hernquist}},\ }\href {\doibase
  10.1093/mnras/stx1887} {\bibfield  {journal} {\bibinfo  {journal} {Mon. Not.
  Roy. Astron. Soc.}\ }\textbf {\bibinfo {volume} {471}},\ \bibinfo {pages}
  {4559} (\bibinfo {year} {2017})},\ \Eprint {http://arxiv.org/abs/1705.05845}
  {arXiv:1705.05845 [astro-ph.CO]} \BibitemShut {NoStop}%
\bibitem [{\citenamefont {Easther}\ \emph {et~al.}(2011)\citenamefont
  {Easther}, \citenamefont {Flauger},\ and\ \citenamefont
  {Gilmore}}]{Easther:2010mr}%
  \BibitemOpen
  \bibfield  {author} {\bibinfo {author} {\bibfnamefont {R.}~\bibnamefont
  {Easther}}, \bibinfo {author} {\bibfnamefont {R.}~\bibnamefont {Flauger}}, \
  and\ \bibinfo {author} {\bibfnamefont {J.~B.}\ \bibnamefont {Gilmore}},\
  }\href {\doibase 10.1088/1475-7516/2011/04/027} {\bibfield  {journal}
  {\bibinfo  {journal} {JCAP}\ }\textbf {\bibinfo {volume} {1104}},\ \bibinfo
  {pages} {027} (\bibinfo {year} {2011})},\ \Eprint
  {http://arxiv.org/abs/1003.3011} {arXiv:1003.3011 [astro-ph.CO]} \BibitemShut
  {NoStop}%
\bibitem [{\citenamefont {Amin}(2010)}]{Amin:2010xe}%
  \BibitemOpen
  \bibfield  {author} {\bibinfo {author} {\bibfnamefont {M.~A.}\ \bibnamefont
  {Amin}},\ }\href@noop {} {\bibfield  {journal} {\bibinfo  {journal} {arXiv
  e-prints}\ } (\bibinfo {year} {2010})},\ \Eprint
  {http://arxiv.org/abs/1006.3075} {arXiv:1006.3075 [astro-ph.CO]} \BibitemShut
  {NoStop}%
\bibitem [{\citenamefont {Amin}\ \emph {et~al.}(2010)\citenamefont {Amin},
  \citenamefont {Easther},\ and\ \citenamefont {Finkel}}]{Amin:2010dc}%
  \BibitemOpen
  \bibfield  {author} {\bibinfo {author} {\bibfnamefont {M.~A.}\ \bibnamefont
  {Amin}}, \bibinfo {author} {\bibfnamefont {R.}~\bibnamefont {Easther}}, \
  and\ \bibinfo {author} {\bibfnamefont {H.}~\bibnamefont {Finkel}},\ }\href
  {\doibase 10.1088/1475-7516/2010/12/001} {\bibfield  {journal} {\bibinfo
  {journal} {JCAP}\ }\textbf {\bibinfo {volume} {1012}},\ \bibinfo {pages}
  {001} (\bibinfo {year} {2010})},\ \Eprint {http://arxiv.org/abs/1009.2505}
  {arXiv:1009.2505 [astro-ph.CO]} \BibitemShut {NoStop}%
\bibitem [{\citenamefont {Schiappacasse}\ and\ \citenamefont
  {Hertzberg}(2018)}]{Schiappacasse:2017ham}%
  \BibitemOpen
  \bibfield  {author} {\bibinfo {author} {\bibfnamefont {E.~D.}\ \bibnamefont
  {Schiappacasse}}\ and\ \bibinfo {author} {\bibfnamefont {M.~P.}\ \bibnamefont
  {Hertzberg}},\ }\href {\doibase 10.1088/1475-7516/2018/03/E01,
  10.1088/1475-7516/2018/01/037} {\bibfield  {journal} {\bibinfo  {journal}
  {JCAP}\ }\textbf {\bibinfo {volume} {1801}},\ \bibinfo {pages} {037}
  (\bibinfo {year} {2018})},\ \bibinfo {note} {[Erratum:
  JCAP1803,no.03,E01(2018)]},\ \Eprint {http://arxiv.org/abs/1710.04729}
  {arXiv:1710.04729 [hep-ph]} \BibitemShut {NoStop}%
\bibitem [{\citenamefont {{Vakhitov}}\ and\ \citenamefont
  {{Kolokolov}}(1973)}]{Vakhitov:1973}%
  \BibitemOpen
  \bibfield  {author} {\bibinfo {author} {\bibfnamefont {N.~G.}\ \bibnamefont
  {{Vakhitov}}}\ and\ \bibinfo {author} {\bibfnamefont {A.~A.}\ \bibnamefont
  {{Kolokolov}}},\ }\href {\doibase 10.1007/BF01031343} {\bibfield  {journal}
  {\bibinfo  {journal} {Radiophysics and Quantum Electronics}\ }\textbf
  {\bibinfo {volume} {16}},\ \bibinfo {pages} {783} (\bibinfo {year}
  {1973})}\BibitemShut {NoStop}%
\bibitem [{\citenamefont {Salmi}\ and\ \citenamefont
  {Hindmarsh}(2012)}]{Salmi:2012ta}%
  \BibitemOpen
  \bibfield  {author} {\bibinfo {author} {\bibfnamefont {P.}~\bibnamefont
  {Salmi}}\ and\ \bibinfo {author} {\bibfnamefont {M.}~\bibnamefont
  {Hindmarsh}},\ }\href {\doibase 10.1103/PhysRevD.85.085033} {\bibfield
  {journal} {\bibinfo  {journal} {Phys. Rev.}\ }\textbf {\bibinfo {volume}
  {D85}},\ \bibinfo {pages} {085033} (\bibinfo {year} {2012})},\ \Eprint
  {http://arxiv.org/abs/1201.1934} {arXiv:1201.1934 [hep-th]} \BibitemShut
  {NoStop}%
\bibitem [{\citenamefont {Hertzberg}(2010)}]{Hertzberg:2010yz}%
  \BibitemOpen
  \bibfield  {author} {\bibinfo {author} {\bibfnamefont {M.~P.}\ \bibnamefont
  {Hertzberg}},\ }\href {\doibase 10.1103/PhysRevD.82.045022} {\bibfield
  {journal} {\bibinfo  {journal} {Phys. Rev.}\ }\textbf {\bibinfo {volume}
  {D82}},\ \bibinfo {pages} {045022} (\bibinfo {year} {2010})},\ \Eprint
  {http://arxiv.org/abs/1003.3459} {arXiv:1003.3459 [hep-th]} \BibitemShut
  {NoStop}%
\bibitem [{\citenamefont {Visinelli}\ \emph {et~al.}(2018)\citenamefont
  {Visinelli}, \citenamefont {Baum}, \citenamefont {Redondo}, \citenamefont
  {Freese},\ and\ \citenamefont {Wilczek}}]{Visinelli:2017ooc}%
  \BibitemOpen
  \bibfield  {author} {\bibinfo {author} {\bibfnamefont {L.}~\bibnamefont
  {Visinelli}}, \bibinfo {author} {\bibfnamefont {S.}~\bibnamefont {Baum}},
  \bibinfo {author} {\bibfnamefont {J.}~\bibnamefont {Redondo}}, \bibinfo
  {author} {\bibfnamefont {K.}~\bibnamefont {Freese}}, \ and\ \bibinfo {author}
  {\bibfnamefont {F.}~\bibnamefont {Wilczek}},\ }\href {\doibase
  10.1016/j.physletb.2017.12.010} {\bibfield  {journal} {\bibinfo  {journal}
  {Phys. Lett.}\ }\textbf {\bibinfo {volume} {B777}},\ \bibinfo {pages} {64}
  (\bibinfo {year} {2018})},\ \Eprint {http://arxiv.org/abs/1710.08910}
  {arXiv:1710.08910 [astro-ph.CO]} \BibitemShut {NoStop}%
\bibitem [{\citenamefont {Manton}(1979)}]{Manton:1978gf}%
  \BibitemOpen
  \bibfield  {author} {\bibinfo {author} {\bibfnamefont {N.~S.}\ \bibnamefont
  {Manton}},\ }\href {\doibase 10.1016/0550-3213(79)90309-2} {\bibfield
  {journal} {\bibinfo  {journal} {Nucl. Phys.}\ }\textbf {\bibinfo {volume}
  {B150}},\ \bibinfo {pages} {397} (\bibinfo {year} {1979})}\BibitemShut
  {NoStop}%
\bibitem [{\citenamefont {{Landy}}\ and\ \citenamefont
  {{Szalay}}(1993)}]{1993ApJ...412...64L}%
  \BibitemOpen
  \bibfield  {author} {\bibinfo {author} {\bibfnamefont {S.~D.}\ \bibnamefont
  {{Landy}}}\ and\ \bibinfo {author} {\bibfnamefont {A.~S.}\ \bibnamefont
  {{Szalay}}},\ }\href {\doibase 10.1086/172900} {\bibfield  {journal}
  {\bibinfo  {journal} {\apj}\ }\textbf {\bibinfo {volume} {412}},\ \bibinfo
  {pages} {64} (\bibinfo {year} {1993})}\BibitemShut {NoStop}%
\bibitem [{\citenamefont {{Wall}}\ and\ \citenamefont
  {{Jenkins}}(2012)}]{2012psa..book.....W}%
  \BibitemOpen
  \bibfield  {author} {\bibinfo {author} {\bibfnamefont {J.~V.}\ \bibnamefont
  {{Wall}}}\ and\ \bibinfo {author} {\bibfnamefont {C.~R.}\ \bibnamefont
  {{Jenkins}}},\ }\href@noop {} {\emph {\bibinfo {title} {Practical Statistics
  for Astronomers, by J.~V.~Wall , C.~R.~Jenkins}}}\ (\bibinfo  {publisher}
  {Cambridge University Press},\ \bibinfo {year} {2012})\BibitemShut {NoStop}%
\bibitem [{\citenamefont {{Saslaw}}(1980)}]{1980ApJ...235..299S}%
  \BibitemOpen
  \bibfield  {author} {\bibinfo {author} {\bibfnamefont {W.~C.}\ \bibnamefont
  {{Saslaw}}},\ }\href {\doibase 10.1086/157634} {\bibfield  {journal}
  {\bibinfo  {journal} {\apj}\ }\textbf {\bibinfo {volume} {235}},\ \bibinfo
  {pages} {299} (\bibinfo {year} {1980})}\BibitemShut {NoStop}%
\bibitem [{\citenamefont {Axenides}\ \emph {et~al.}(2000)\citenamefont
  {Axenides}, \citenamefont {Komineas}, \citenamefont {Perivolaropoulos},\ and\
  \citenamefont {Floratos}}]{Axenides:1999hs}%
  \BibitemOpen
  \bibfield  {author} {\bibinfo {author} {\bibfnamefont {M.}~\bibnamefont
  {Axenides}}, \bibinfo {author} {\bibfnamefont {S.}~\bibnamefont {Komineas}},
  \bibinfo {author} {\bibfnamefont {L.}~\bibnamefont {Perivolaropoulos}}, \
  and\ \bibinfo {author} {\bibfnamefont {M.}~\bibnamefont {Floratos}},\ }\href
  {\doibase 10.1103/PhysRevD.61.085006} {\bibfield  {journal} {\bibinfo
  {journal} {Phys. Rev.}\ }\textbf {\bibinfo {volume} {D61}},\ \bibinfo {pages}
  {085006} (\bibinfo {year} {2000})},\ \Eprint
  {http://arxiv.org/abs/hep-ph/9910388} {arXiv:hep-ph/9910388 [hep-ph]}
  \BibitemShut {NoStop}%
\bibitem [{\citenamefont {Battye}\ and\ \citenamefont
  {Sutcliffe}(2000)}]{Battye:2000qj}%
  \BibitemOpen
  \bibfield  {author} {\bibinfo {author} {\bibfnamefont {R.}~\bibnamefont
  {Battye}}\ and\ \bibinfo {author} {\bibfnamefont {P.}~\bibnamefont
  {Sutcliffe}},\ }\href {\doibase 10.1016/S0550-3213(00)00506-X} {\bibfield
  {journal} {\bibinfo  {journal} {Nucl. Phys.}\ }\textbf {\bibinfo {volume}
  {B590}},\ \bibinfo {pages} {329} (\bibinfo {year} {2000})},\ \Eprint
  {http://arxiv.org/abs/hep-th/0003252} {arXiv:hep-th/0003252 [hep-th]}
  \BibitemShut {NoStop}%
\bibitem [{\citenamefont {{Aitchison}}\ \emph {et~al.}(1991)\citenamefont
  {{Aitchison}}, \citenamefont {{Weiner}}, \citenamefont {{Silberberg (Bell)}},
  \citenamefont {{Leaird}}, \citenamefont {{Oliver}}, \citenamefont
  {{Jackel}},\ and\ \citenamefont {{Smith}}}]{Aitchison:91}%
  \BibitemOpen
  \bibfield  {author} {\bibinfo {author} {\bibfnamefont {J.~S.}\ \bibnamefont
  {{Aitchison}}}, \bibinfo {author} {\bibfnamefont {A.~M.}\ \bibnamefont
  {{Weiner}}}, \bibinfo {author} {\bibfnamefont {Y.}~\bibnamefont {{Silberberg
  (Bell)}}}, \bibinfo {author} {\bibfnamefont {D.~E.}\ \bibnamefont
  {{Leaird}}}, \bibinfo {author} {\bibfnamefont {M.~K.}\ \bibnamefont
  {{Oliver}}}, \bibinfo {author} {\bibfnamefont {J.~L.}\ \bibnamefont
  {{Jackel}}}, \ and\ \bibinfo {author} {\bibfnamefont {P.~W.~E.}\ \bibnamefont
  {{Smith}}},\ }\href {\doibase 10.1364/OL.16.000015} {\bibfield  {journal}
  {\bibinfo  {journal} {Optics Letters}\ }\textbf {\bibinfo {volume} {16}},\
  \bibinfo {pages} {15} (\bibinfo {year} {1991})}\BibitemShut {NoStop}%
\bibitem [{\citenamefont {{Nguyen}}\ \emph {et~al.}(2017)\citenamefont
  {{Nguyen}}, \citenamefont {{Luo}},\ and\ \citenamefont
  {{Hulet}}}]{2017Sci...356..422N}%
  \BibitemOpen
  \bibfield  {author} {\bibinfo {author} {\bibfnamefont {J.~H.~V.}\
  \bibnamefont {{Nguyen}}}, \bibinfo {author} {\bibfnamefont {D.}~\bibnamefont
  {{Luo}}}, \ and\ \bibinfo {author} {\bibfnamefont {R.~G.}\ \bibnamefont
  {{Hulet}}},\ }\href {\doibase 10.1126/science.aal3220} {\bibfield  {journal}
  {\bibinfo  {journal} {Science}\ }\textbf {\bibinfo {volume} {356}},\ \bibinfo
  {pages} {422} (\bibinfo {year} {2017})},\ \Eprint
  {http://arxiv.org/abs/1703.04662} {arXiv:1703.04662 [cond-mat.quant-gas]}
  \BibitemShut {NoStop}%
\bibitem [{\citenamefont {Cotner}(2016)}]{Cotner:2016aaq}%
  \BibitemOpen
  \bibfield  {author} {\bibinfo {author} {\bibfnamefont {E.}~\bibnamefont
  {Cotner}},\ }\href {\doibase 10.1103/PhysRevD.94.063503} {\bibfield
  {journal} {\bibinfo  {journal} {Phys. Rev.}\ }\textbf {\bibinfo {volume}
  {D94}},\ \bibinfo {pages} {063503} (\bibinfo {year} {2016})},\ \Eprint
  {http://arxiv.org/abs/1608.00547} {arXiv:1608.00547 [astro-ph.CO]}
  \BibitemShut {NoStop}%
\bibitem [{\citenamefont {Helfer}\ \emph {et~al.}(2017)\citenamefont {Helfer},
  \citenamefont {Marsh}, \citenamefont {Clough}, \citenamefont {Fairbairn},
  \citenamefont {Lim},\ and\ \citenamefont {Becerril}}]{Helfer:2016ljl}%
  \BibitemOpen
  \bibfield  {author} {\bibinfo {author} {\bibfnamefont {T.}~\bibnamefont
  {Helfer}}, \bibinfo {author} {\bibfnamefont {D.~J.~E.}\ \bibnamefont
  {Marsh}}, \bibinfo {author} {\bibfnamefont {K.}~\bibnamefont {Clough}},
  \bibinfo {author} {\bibfnamefont {M.}~\bibnamefont {Fairbairn}}, \bibinfo
  {author} {\bibfnamefont {E.~A.}\ \bibnamefont {Lim}}, \ and\ \bibinfo
  {author} {\bibfnamefont {R.}~\bibnamefont {Becerril}},\ }\href {\doibase
  10.1088/1475-7516/2017/03/055} {\bibfield  {journal} {\bibinfo  {journal}
  {JCAP}\ }\textbf {\bibinfo {volume} {1703}},\ \bibinfo {pages} {055}
  (\bibinfo {year} {2017})},\ \Eprint {http://arxiv.org/abs/1609.04724}
  {arXiv:1609.04724 [astro-ph.CO]} \BibitemShut {NoStop}%
\bibitem [{\citenamefont {Widdicombe}\ \emph {et~al.}(2018)\citenamefont
  {Widdicombe}, \citenamefont {Helfer}, \citenamefont {Marsh},\ and\
  \citenamefont {Lim}}]{Widdicombe:2018oeo}%
  \BibitemOpen
  \bibfield  {author} {\bibinfo {author} {\bibfnamefont {J.~Y.}\ \bibnamefont
  {Widdicombe}}, \bibinfo {author} {\bibfnamefont {T.}~\bibnamefont {Helfer}},
  \bibinfo {author} {\bibfnamefont {D.~J.~E.}\ \bibnamefont {Marsh}}, \ and\
  \bibinfo {author} {\bibfnamefont {E.~A.}\ \bibnamefont {Lim}},\ }\href
  {\doibase 10.1088/1475-7516/2018/10/005} {\bibfield  {journal} {\bibinfo
  {journal} {JCAP}\ }\textbf {\bibinfo {volume} {1810}},\ \bibinfo {pages}
  {005} (\bibinfo {year} {2018})},\ \Eprint {http://arxiv.org/abs/1806.09367}
  {arXiv:1806.09367 [astro-ph.CO]} \BibitemShut {NoStop}%
\bibitem [{\citenamefont {Hogan}\ and\ \citenamefont
  {Rees}(1988)}]{Hogan:1988mp}%
  \BibitemOpen
  \bibfield  {author} {\bibinfo {author} {\bibfnamefont {C.~J.}\ \bibnamefont
  {Hogan}}\ and\ \bibinfo {author} {\bibfnamefont {M.~J.}\ \bibnamefont
  {Rees}},\ }\href {\doibase 10.1016/0370-2693(88)91655-3} {\bibfield
  {journal} {\bibinfo  {journal} {Phys. Lett.}\ }\textbf {\bibinfo {volume}
  {B205}},\ \bibinfo {pages} {228} (\bibinfo {year} {1988})}\BibitemShut
  {NoStop}%
\bibitem [{\citenamefont {Kallosh}\ and\ \citenamefont
  {Linde}(2013{\natexlab{a}})}]{Kallosh:2013xya}%
  \BibitemOpen
  \bibfield  {author} {\bibinfo {author} {\bibfnamefont {R.}~\bibnamefont
  {Kallosh}}\ and\ \bibinfo {author} {\bibfnamefont {A.}~\bibnamefont
  {Linde}},\ }\href {\doibase 10.1088/1475-7516/2013/06/028} {\bibfield
  {journal} {\bibinfo  {journal} {JCAP}\ }\textbf {\bibinfo {volume} {1306}},\
  \bibinfo {pages} {028} (\bibinfo {year} {2013}{\natexlab{a}})},\ \Eprint
  {http://arxiv.org/abs/1306.3214} {arXiv:1306.3214 [hep-th]} \BibitemShut
  {NoStop}%
\bibitem [{\citenamefont {Kallosh}\ and\ \citenamefont
  {Linde}(2013{\natexlab{b}})}]{Kallosh:2013hoa}%
  \BibitemOpen
  \bibfield  {author} {\bibinfo {author} {\bibfnamefont {R.}~\bibnamefont
  {Kallosh}}\ and\ \bibinfo {author} {\bibfnamefont {A.}~\bibnamefont
  {Linde}},\ }\href {\doibase 10.1088/1475-7516/2013/07/002} {\bibfield
  {journal} {\bibinfo  {journal} {JCAP}\ }\textbf {\bibinfo {volume} {1307}},\
  \bibinfo {pages} {002} (\bibinfo {year} {2013}{\natexlab{b}})},\ \Eprint
  {http://arxiv.org/abs/1306.5220} {arXiv:1306.5220 [hep-th]} \BibitemShut
  {NoStop}%
\bibitem [{\citenamefont {Frolov}(2008)}]{Frolov:2008hy}%
  \BibitemOpen
  \bibfield  {author} {\bibinfo {author} {\bibfnamefont {A.~V.}\ \bibnamefont
  {Frolov}},\ }\href {\doibase 10.1088/1475-7516/2008/11/009} {\bibfield
  {journal} {\bibinfo  {journal} {JCAP}\ }\textbf {\bibinfo {volume} {0811}},\
  \bibinfo {pages} {009} (\bibinfo {year} {2008})},\ \Eprint
  {http://arxiv.org/abs/0809.4904} {arXiv:0809.4904 [hep-ph]} \BibitemShut
  {NoStop}%
\bibitem [{\citenamefont {Fodor}\ \emph {et~al.}(2008)\citenamefont {Fodor},
  \citenamefont {Forgacs}, \citenamefont {Horvath},\ and\ \citenamefont
  {Lukacs}}]{Fodor:2008es}%
  \BibitemOpen
  \bibfield  {author} {\bibinfo {author} {\bibfnamefont {G.}~\bibnamefont
  {Fodor}}, \bibinfo {author} {\bibfnamefont {P.}~\bibnamefont {Forgacs}},
  \bibinfo {author} {\bibfnamefont {Z.}~\bibnamefont {Horvath}}, \ and\
  \bibinfo {author} {\bibfnamefont {A.}~\bibnamefont {Lukacs}},\ }\href
  {\doibase 10.1103/PhysRevD.78.025003} {\bibfield  {journal} {\bibinfo
  {journal} {Phys. Rev.}\ }\textbf {\bibinfo {volume} {D78}},\ \bibinfo {pages}
  {025003} (\bibinfo {year} {2008})},\ \Eprint {http://arxiv.org/abs/0802.3525}
  {arXiv:0802.3525 [hep-th]} \BibitemShut {NoStop}%
\bibitem [{\citenamefont {Mukaida}\ \emph {et~al.}(2017)\citenamefont
  {Mukaida}, \citenamefont {Takimoto},\ and\ \citenamefont
  {Yamada}}]{Mukaida:2016hwd}%
  \BibitemOpen
  \bibfield  {author} {\bibinfo {author} {\bibfnamefont {K.}~\bibnamefont
  {Mukaida}}, \bibinfo {author} {\bibfnamefont {M.}~\bibnamefont {Takimoto}}, \
  and\ \bibinfo {author} {\bibfnamefont {M.}~\bibnamefont {Yamada}},\ }\href
  {\doibase 10.1007/JHEP03(2017)122} {\bibfield  {journal} {\bibinfo  {journal}
  {JHEP}\ }\textbf {\bibinfo {volume} {03}},\ \bibinfo {pages} {122} (\bibinfo
  {year} {2017})},\ \Eprint {http://arxiv.org/abs/1612.07750} {arXiv:1612.07750
  [hep-ph]} \BibitemShut {NoStop}%
\bibitem [{\citenamefont {Saffin}\ \emph {et~al.}(2014)\citenamefont {Saffin},
  \citenamefont {Tognarelli},\ and\ \citenamefont {Tranberg}}]{Saffin:2014yka}%
  \BibitemOpen
  \bibfield  {author} {\bibinfo {author} {\bibfnamefont {P.~M.}\ \bibnamefont
  {Saffin}}, \bibinfo {author} {\bibfnamefont {P.}~\bibnamefont {Tognarelli}},
  \ and\ \bibinfo {author} {\bibfnamefont {A.}~\bibnamefont {Tranberg}},\
  }\href {\doibase 10.1007/JHEP08(2014)125} {\bibfield  {journal} {\bibinfo
  {journal} {JHEP}\ }\textbf {\bibinfo {volume} {08}},\ \bibinfo {pages} {125}
  (\bibinfo {year} {2014})},\ \Eprint {http://arxiv.org/abs/1401.6168}
  {arXiv:1401.6168 [hep-ph]} \BibitemShut {NoStop}%
\end{thebibliography}%

\newpage
\onecolumngrid
\section{Appendix}

\subsection{Connection to a Relativistic Model}
\label{sec:RelativisticNR}

In the main body of the paper we did not include a detailed analysis of the non-relativistic limit of strongly self-interacting relativistic theories (if it exists). We took certain non-relativistic field equations with strong self-interactions as given, and explored the solutions. As we discuss below, the equations we use can be   justified as being obtained by integrating out  the fast time variation in the weakly interacting limit. While we believe that some aspects of the relativistic -- non-relativistic connection persists at large self-interactions as well, a rigorous mapping is beyond the scope of the present work. We note that even with strong self interactions, the spatio-temporal variations of the solutions of the system under consideration remain non-relativistic, and the gravitational potential remains small, making the exploration in the main body of the paper self-consistent in this respect. 

To derive our equations of motion \eqref{eq:MasterEq} from a relativistic scalar field theory (in a particular limit discussed below), consider a real scalar field $\phi$ within general relativity.  Consider a real scalar field minimally coupled to gravity with the action
\beq
S = \int \frac{d^4 x}{\hbar c^2}\sqrt{-g} \left[\frac{R}{16\pi G}-\frac{1}{2} g^{\mu\nu} \partial_{\nu}\phi \partial_{\mu} \phi  - V(\phi) \right]\,,
\eeq
where $\phi$ has dimensions of energy, $R$ is the Ricci scalar, $g_{\mu\nu}$ is the metric, $g$ is the determinant of the metric, and $d^4 x=(cdt)d^3x$.  We are interested in potentials of the form
\beq
V(\phi)= \frac{m^2c^2}{2\hbar^2}\phi^2+V_{\rm nl}(\phi) \,,
\eeq
where $V_{\rm nl}(\phi)$ contains the non-quadratic part of the potential, whose shape is controlled by a scale $M$.  As a concrete example, we can consider the potential $V(\phi)=(m^2 M^2/2)\tanh^2(\phi/M)$ \cite{Kallosh:2013xya,Kallosh:2013hoa}, although the precise form is not necessary for most of the discussion that follows.

\noindent\subsubsection{The weak field approximation -- non-expanding spacetime}

In the weak field limit (i.e. for $\Phi/ c^2 \ll 1$ where $\Phi$ is the Newtonian gravitational potential) and in the absence of expansion, the metric is determined by the line element of the form
\beq\label{ds2}
ds^2 = \left(1+2\frac{\Phi}{c^2}\right) (cdt)^2 -  \left(1-2\frac{\Phi}{c^2}\right) d{\bm x}^2\,.
\eeq
Note that we are ignoring anisotropic stress, as well as vector and tensor perturbations. In the linear regime, anisotropic stress is absent and will be absent away from solitons at the very least. In a time averaged sense,  the anisotropic stress will be small inside the solitons.

The equation of motion (nonlinear Klein-Gordon equation in curved spacetime) satisfied by the field $\phi$ is
\beq
\frac{1}{\sqrt{-g}}\partial_{\mu}\left(\sqrt{-g}\,g^{\mu\nu}\partial_{\nu}\phi\right) + \partial_{\phi}V(\phi) = 0\,,
\eeq
which, to leading order in $\Phi/c^2$, yields
\beq\label{eomfull1}
\frac{\partial^2\phi}{\partial (ct)^2} - \left(1+4\frac{\Phi}{c^2}\right)\nabla^2\phi  - \frac{4}{c^2}\frac{\partial \Phi}{\partial (ct)}\frac{\partial\phi}{\partial (ct)} + \left(1+2\frac{\Phi}{c^2}\right)\partial_{\phi}V=0\,.
\eeq
In turn, the Einstein equations reduce to the Poisson equation
\beq\label{eomfull2}
\nabla^2\Phi = \frac{4\pi G}{c^2}T^0{}_0\,,
\eeq
with
\beq\label{T00}
T^0{}_0 = \frac{1}{2}\left(1-2\frac{\Phi}{c^2}\right)\left(\frac{\partial\phi}{\partial(ct)}\right)^2 + \frac{1}{2}\left(1+2\frac{\Phi}{c^2}\right)(\nabla\phi)^2 + V(\phi)\,.
\eeq

\subsubsection{The non-relativistic limit}

In order to consider the ``non-relativistic'' limit, it is convenient to redefine the real scalar $\phi$ in terms of a complex field $\psi$, factoring out the rest energy contribution
\beq
\phi = \frac{\hbar}{\sqrt{2}\, m} \left(\psi e^{-imc^2t/\hbar} + {\rm h.c.}\right)=\sqrt{2}\frac{\hbar}{m}\Re[\psi e^{-imc^2t/\hbar}] \,,
\eeq
where the normalization constant is chosen so that (\ref{eomfull2}) reduces to the usual non-relativistic Poisson form. 
Straightforward substitution into (\ref{eomfull1}) and (\ref{T00}) yields
\Beq
\Bigg[ i\hbar\frac{\partial \psi}{\partial t} &- \frac{\hbar^2}{2mc^2}\frac{\partial^2\psi}{\partial t^2}  + \frac{\hbar^2}{2m}\left(1+4\frac{\Phi}{c^2}\right)\nabla^2\psi + \frac{2\hbar^2}{mc^4}\frac{\partial \Phi}{\partial t} \left(\frac{\partial \psi}{\partial t} - \frac{imc^2}{\hbar}\psi\right)  + \frac{1}{2} mc^2 \psi \Bigg] e^{-imc^2t/\hbar}  +  ({\rm h.c.}) \\
&- \left(1+2\frac{\Phi}{c^2}\right) \left[\frac{mc^2}{2}(\psi e^{-i mc^2t/\hbar}+{\rm .h.c})+\frac{\hbar}{\sqrt{2}}\partial_{\phi}V_{\rm nl}(\phi)\right] = 0\,, \label{eompsi1}
\Eeq
and
\begin{align}\label{T00psi}
T_0^0 &= \frac{\hbar^2}{4m^2c^2}\left(1-2\frac{\Phi}{c^2}\right)\left[\left(\frac{\partial\psi}{\partial t} - \frac{imc^2}{\hbar}\psi\right)e^{-imc^2t/\hbar} + {\rm h.c.}\right]^2\\
&\quad  + \frac{\hbar^2}{4m^2}\left(1+2\frac{\Phi}{c^2}\right)\left[\left(\nabla \psi\right)e^{-imc^2t/\hbar} + {\rm h.c.}\right]^2 + \frac{c^2}{4}(\psi e^{-imc^2t/\hbar} + {\rm h.c.})^2+V_{\rm nl}.\notag
\end{align}

Let $\tau_m=\hbar/mc^2$ and $\lambda_m=\hbar/mc$. Now let us assume that $|\tau_m\partial_t\psi|\ll |\psi|$; similarly $|\tau_m\partial_t\Phi|\ll |\Phi|$.  We now average $T^0_0$ over a period $2\pi \tau_m$ assuming that $\Phi$ and $\psi$ do not change appreciably over this period.\footnote{This part is not entirely rigorous, and it deserves to be handled with care.}  This yields

\beq
\langle T_0^0\rangle = \frac{c^2}{2}\left(1-2\frac{\Phi}{c^2}\right)\left| \tau_m\frac{\partial\psi}{\partial t} - i\psi\right|^2 + \frac{1}{2}\left(1+2\frac{\Phi}{c^2}\right)c^2\lambda_m^2\left|\nabla \psi\right|^2 + \frac{c^2}{2}|\psi|^2+ \langle V_{\rm nl}\rangle\,.
\eeq
It is convenient to define 
\Beq
\label{eq:VUconnect}
\langle V_{\rm nl}\rangle \equiv U_{\rm nl}(|\psi|^2)\,.
\Eeq
 Assuming $|\Phi/c^2|\ll 1$, and $|\tau_m\partial_t\psi|\ll |\psi|$, the above expression simplifies to 
\beq
\langle T^0{}_0\rangle = c^2|\psi|^2 + \frac{1}{2}c^2\lambda_m^2\left|\nabla \psi\right|^2+ U_{\rm nl}+\mathcal{O}[\Phi/c^2,\tau_m\partial_t]\,.
\eeq

To get the Schr\"odinger-like equation, we multiply eq.~\eqref{eompsi1} by $e^{it/\tau_m}$ and average over a period $2\pi\tau_m$, again assuming that $\Phi$ and $\psi$ do not change appreciably over this period. This temporal averaging will get rid of the h.c part in \eqref{eompsi1}. Moreover, we divide the resulting equation by $mc^2$, to get
\Beq
\Bigg[ i\tau_m\frac{\partial \psi}{\partial t} &- \frac{1}{2}\tau_m^2\frac{\partial^2\psi}{\partial t^2}  + \left(1+4\frac{\Phi}{c^2}\right)\frac{1}{2}\lambda_m^2\nabla^2\psi + 2\tau_m\frac{\partial (\Phi/c^2)}{\partial t} \left(\tau_m\frac{\partial \psi}{\partial t} - i\psi\right) + \frac{1}{2} mc^2 \psi \Bigg] \\ \label{KGpsiAvg}
 &  - \left(1+2\frac{\Phi}{c^2}\right) \frac{\tau_m}{\sqrt{2}}\langle e^{it/\tau_m}\partial_{\phi}V\rangle = 0\,.
\Eeq
We will show in the next subsection that $\frac{\tau_m}{\sqrt{2}}\langle e^{it/\tau_m}\partial_{\phi}V\rangle=\psi\partial_{|\psi|^2}U_{\rm nl}(|\psi|)^2=\psi U_{\rm nl}'(|\psi|^2)$. 

Treating $\Phi/c^2$ and $\tau_m\partial_t$ as separate small parameters, and keeping leading order terms in each (but ignoring $\Phi/c^2\times\tau_m\partial_t$), we have
\Beq
\label{KGpsiAvg1}
i\tau_m\frac{\partial \psi}{\partial t}  + \left(1+4\frac{\Phi}{c^2}\right)\frac{1}{2}\lambda_m^2\nabla^2\psi -\frac{\Phi}{c^2}\psi-\left(1+2\frac{\Phi}{c^2}\right)\psi U_{\rm nl}'(|\psi|^2)= 0\,.
\Eeq
After some re-arranging
\Beq
\label{KGpsiAvg2}
i\tau_m\frac{\partial \psi}{\partial t}+\frac{1}{2}\lambda_m^2\nabla^2\psi -\psi U_{\rm nl}'(|\psi|^2) -\frac{\Phi}{c^2}\left(\psi-2\lambda_m^2\nabla^2\psi+2\psi U_{\rm nl}'(|\psi|^2)\right) = 0\,.
\Eeq
On the one hand it is clear that $\nabla^2\psi\gg (\Phi/c^2)\nabla^2\psi$. However, for large amplitude solitons $\lambda_m^2\nabla^2\psi\sim \psi$; hence, it is not clear that we can drop this term compared to the $\psi$ in the term with the $\Phi$ coefficient (also see \cite{Namjoo:2017nia}). A similar argument holds for $\langle \hdots\rangle$ terms. For the discussion that follows, we will use $\nabla^2\psi\gg (\Phi/c^2)\nabla^2\psi$, to arrive at

\Beq
\label{KGpsiAvg3}
i\tau_m\frac{\partial \psi}{\partial t}+\frac{1}{2}\lambda_m^2\nabla^2\psi -\psi U_{\rm nl}'(|\psi|^2) -\frac{\Phi}{c^2}\psi = 0\,.
\Eeq
Using our result for $\langle T^0_0\rangle$, we also have the Poisson equation at the lowest order in $\Phi/c^2$
\Beq
\nabla^2\Phi=4\pi G \left[|\psi|^2+\frac{1}{2}\lambda_m^2\left|\nabla \psi\right|^2+ c^{-2}U_{\rm nl}(|\psi|^2)\right]\,.
\Eeq
These are our master equations used for time evolution of the field and for determining the metric potential (see eq.~\eqref{eq:MasterEq}).
In arriving at eq.~\eqref{eq:MasterEq} in this limit (ignoring expansion for the moment), we  assumed weak field gravity and restricted ourselves to scalar metric perturbations without anisotropic stress. 

Since we were not interested in reproducing the limit of a particular relativistic theory in the main body of the text, we simply took $U_{\rm nl}$ to be an effective potential for our theory. Nevertheless, by using \eqref{eq:VUconnect} we can link $U_{\rm nl}$ to $V_{\rm nl}$ at least for small amplitudes. We turn to this task next.

The time averaging procedure in eq.~\eqref{eq:VUconnect} is mathematically well defined  for any potential which admits a Taylor expansion and has a quadratic minimum. 
\Beq
V(\phi)&=m^2 M^2\sum_{n=1}^{\infty}a_n \left(\frac{\phi}{M}\right)^{2n}\,,\qquad &\textrm{where}& \qquad a_1=1/2\,.
\Eeq
The non-quadratic ({\it nonlinear} part) of this potential is
\Beq
V_{\rm nl}&=V(\phi)-\frac{1}{2}m^2\phi^2\,.
\Eeq
Using $\phi=\sqrt{2}\Re[\psi e^{-it/\tau_m}]$ and taking a time average of this nonlinear part over a period $2\pi\tau_m$, we have
\Beq
U_{\rm nl}(|\psi|^2)\equiv\langle V_{\rm nl}\rangle&=m^2M^2\sum_{n=2}^{\infty}b_n \left(\frac{|\psi|^2}{m^2M^2}\right)^{n} \qquad &\textrm{where}& \qquad b_n=\frac{(2n)!}{2^n(n!)^2}\,a_n\,.
\Eeq
We also need the time average of $\partial_\phi V_{\rm nl}$ for the equations of motion:
\Beq
\psi U_{\rm nl}'(|\psi|^2)=\frac{\tau_m}{\sqrt{2}}\langle e^{it/\tau_m}\partial_{\phi}V_{\rm nl}\rangle&=\psi\sum_{n=2}^{\infty}c_n \psi\left(\frac{|\psi|^2}{m^2M^2}\right)^{n-1} \qquad &\textrm{where}& \qquad c_n=\frac{(2n-1)!}{2^{n-1}(n-1!)^2}\,a_n=\frac{b_n}{n}
\Eeq
It is beneficial to have a fitting function for $\langle V_{\rm nl}\rangle$. For a potential of the form
\Beq
V(\phi)=\frac{m^2M^2}{2}\tanh^2\left(\frac{\phi}{M}\right)\,,
\Eeq
we can find that for $|\psi|\le \pi/(2\sqrt{2})$, an excellent approximation to $U_{\rm nl}(|\psi|^2)$ is provided by \Beq
U_{\rm nl}(|\psi|^2)=-\frac{|\psi|^2}{2}\dfrac{\dfrac{|\psi|^2}{m^2M^2}}{1+\dfrac{|\psi|^2}{m^2M^2}}\,,\qquad\textrm{and correspondingly}\qquad U_{\rm nl}'(|\psi|^2)=-\frac{|\psi|^2}{m^2M^2}\dfrac{1+\dfrac{|\psi|^2}{2m^2M^2}}{\left(1+\dfrac{|\psi|^2}{m^2M^2}\right)^2}\,.
\Eeq
Rescaling our field by $mM$, we recover the potential used in the main body of the text. We caution, that the form beyond $|\psi| > \pi/(2\sqrt{2})$ need not be simply connected to the relativistic potential. Moreover, at these large amplitudes, we might benefit by time-averaging over amplitude-dependent frequencies.

To include the effect of background expansion we consider a metric of the form 
\Beq
ds^2=(1+2\Phi)(cdt^2)-a^2(t)(1-2\Phi)d{\bm x}^2\,.
\Eeq
where $a(t)$ is the scalefactor. Our complete set of equations then becomes (under the assumption that $H^{-1}\gg \tau_m$ and $c/H\gg \lambda_m$),
\Beq
&\left[i\left(\partial_t+\frac{3}{2}H\right)+\frac{1}{2a^2}\nabla^2 -U_{\rm nl}'(|\psi|^2) -\Phi\right]\psi=0\,,\\
&\frac{\nabla^2}{a^2}\Phi=\frac{\beta^2}{2}\left[|\psi|^2+\frac{1}{2a^2}|\nabla\psi|^2+U_{\rm nl}(|\psi|^2)\right]-\frac{3}{2}H^2\,,\\
&H^2=\frac{\beta^2}{3}\overline{\left[|\psi|^2+\frac{1}{2a^2}|\nabla\psi|^2+U_{\rm nl}(|\psi|^2)\right]}\,,
\Eeq
where $\overline{[\hdots]}$ indicates a spatial average. The third equation is obtained from the Einstein equations for a homogeneous and isotropic universe (the Friedmann equation).  This completes our derivation, with caveats, of the {\it master equations} \eqref{eq:MasterEq}  that are used in the main body of the paper.

For recent derivations and discussions of the non-relativistic limit, as well as decay rates for solitons with and without weak-field gravity (but in a non-expanding universe), see \cite{Namjoo:2017nia,Eby:2018ufi}.

\subsection{Details on Initial Conditions}
\label{sec:InitialConditions}
In this appendix, we derive the vacuum initial conditions for a free non-relativistic field from the appropriate relativistic free field vacuum perturbations. Starting with the definition of the Fourier transform:
\Beq
\frac{1}{\sqrt{V}} \sum_\bk e^{-i\bk\cdot \bx}\phi_{\bk}(t)
&=\phi(t,\bx)\,,\\
&=\frac{1}{\sqrt{2}m}\left[\psi(t,\bx)e^{-imt}+\rm{c.c}\right]\,,\\
&=\frac{1}{\sqrt{2}m}\left[\{\psi^R(t,\bx)+i\psi^I(t,\bx)\}e^{-imt}+\rm{c.c}\right]\,,\\
&=\frac{\sqrt{2}}{m}\left[\psi^R(t,\bx)\cos(mt)+\psi^I(t,\bx)\sin (mt)\right]\,,\\
&=\frac{1}{\sqrt{V}}\sum_\bk \frac{\sqrt{2}}{m}e^{-i\bk\cdot\bx}\left[\psi^R_\bk(t)\cos(mt)+\psi^I_\bk(t)\sin (mt)\right]\,,
\Eeq
where $\psi^{R,I}_\bk(t)$ are the Fourier transforms of $\psi^{R,I}(t,\bx)$. Hence we have
\Beq
\psi^R_\bk(t)\cos(mt)+\psi^I_\bk(t)\sin (mt)=\frac{m}{\sqrt{2}}\phi_\bk(t)\,.
\Eeq
Similarly we have
\Beq
-\psi^R_\bk(t)\sin(mt)+\psi^I_\bk(t)\cos (mt)\approx\frac{1}{\sqrt{2}}\dot{\phi}_\bk(t)\,,
\Eeq
where we have assumed $|\dot{\psi}^{R,I}_\bk(t)|/m\ll |\psi^{I,R}_\bk(t)|$. At an initial time $t=0$, we have
\Beq
\psi^R_\bk(0)=\frac{m}{\sqrt{2}}\phi_\bk(0)\,\qquad\textrm{and}\qquad \psi^I_\bk(0)\approx\frac{1}{\sqrt{2}}\dot{\phi}_\bk(0)\,.
\Eeq
Now, following the implementation in Defrost \cite{Frolov:2008hy}, we can write $\phi_\bk(0)=b_\bk/\sqrt{2\omega_k}$ and $\dot{\phi}_\bk(0)=c_\bk\sqrt{\omega_k/2}$ where $\langle b_\bk b_{\bq}^*\rangle =\delta_{\bk\bq}$ and $\langle c_\bk c_{\bq}^*\rangle =\delta_{\bk\bq}$ with $b_\bk$ and $c_\bk$ independent complex numbers (4 identically distributed independent variables). The real and imaginary parts of each are drawn from a zero mean Gaussian distribution, with a variance of $1/2$. Then, we have
\Beq
\psi^R_\bk(0)=\frac{m}{2\sqrt{\omega_k}}b_\bk \approx \frac{\sqrt{m}}{2}b_\bk \,\qquad \textrm{and}\qquad\psi^I_\bk(0)\approx\frac{\sqrt{\omega_k}}{2}c_\bk\approx\frac{\sqrt{m}}{2}c_\bk\,,
\Eeq
where in the second equality we assumed $k\ll m$ so that we have $\omega_k=\sqrt{k^2+m^2}\approx m$. 

We are interested in $\psi_\bk(t)$ which is the Fourier transform of $\psi(t,\bx)$. It can be written as 
\Beq
\psi_\bk(0)=\psi^R_\bk(0)+i\psi^I_\bk(0)\approx \frac{\sqrt{m}}{2}\left\{(\Re[b_\bk]-\Im[c_\bk])+i(\Im[b_\bk]+\Re[c_\bk])\right\}\,.
\Eeq
Hence, 
\Beq
\langle|\psi_\bk(0)|^2\rangle\approx\frac{m}{2}\,,
\Eeq
with amplitude drawn from a Raleigh distribution and the phase drawn from a  uniform distribution. This is consistent with the result in the main body of the paper. 

\begin{figure}[t!] 
   \centering
   \includegraphics[width=6in]{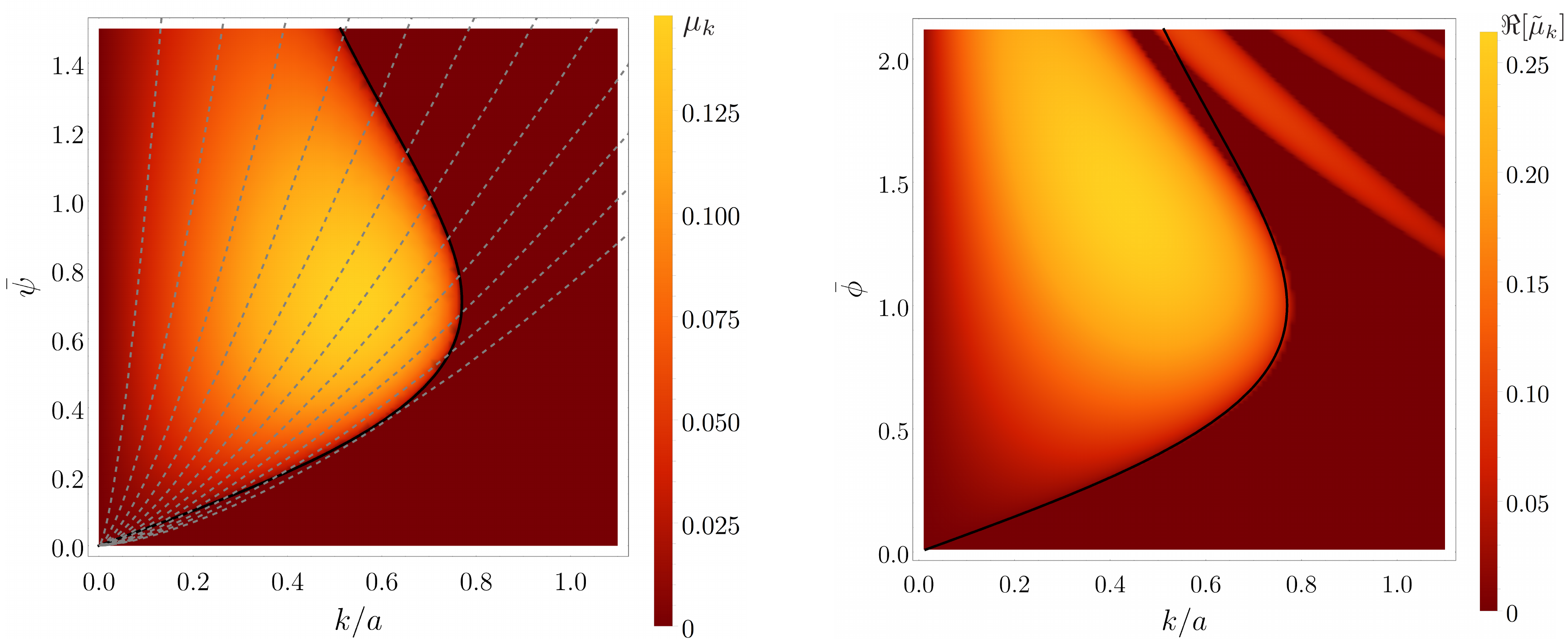} 
   \caption{Left: Colors show the growth rate $\mu_k$ as a function of $k$ and $\bar{\psi}$. The dark red regions are stable. The color bar indicates the magnitude of the $\mu_k$ and $\tilde{\mu}_k$. The dotted lines indicate the flow of $k$ and $\bar{\psi}$ as the universe expands. To compare this plot with the corresponding Floquet chart (right) from the relativistic case, we set $\bar{\psi}=\bar{\phi}/\sqrt{2}$. The factor of $\sqrt{2}$ can be seen from $\phi=\sqrt{2}\Re[\psi e^{-it}]$. The magnitude of the growth rate of the instability and the boundary of the non-relativistic instability band (solid black line) deviate from the relativistic one at large amplitudes. The same is true (to a larger extent) for the magnitude of the Floquet exponent. Also notice that the higher order instability bands are absent in the non-relativistic treatment. We use $V_{\rm nl}(\phi)=(1/2)m^2M^2\tanh^2(\phi/M)-(1/2)m^2\phi^2$ and $U_{\rm nl}(|\psi|^2)= \langle V_{\rm nl}(\phi)\rangle\approx-|\psi|^4/2(1+|\psi|^2)$ for $\phi/M<\pi/2$, and the comparison at large $\phi$, $\psi$ is not justified.}
   \label{fig:LINR}
\end{figure}
\subsection{Comparison of Linear Instability Relativistic and non-relativistic Systems}
The instability analysis discussed in the main text is connected to Floquet analysis in the corresponding relativistic theory (see for example, \cite{Lozanov:2017hjm}). However, the instability bands as well as the Floquet exponents can differ from the relativistic case at large amplitudes and relativistic wave-numbers. For the relativistic version (with $a=1$, $H=0$), the perturbation to the homogeneous field satisfies:
\Beq
\partial_t^2\delta\phi_\bk+\left[k^2+1+V_{\rm nl}''(\bar{\phi})\right]\delta\phi_\bk=0\,,
\Eeq
where the field $\phi$ is measured in units of $M$ and spacetime in units of $m^{-1}$. The periodic term in $V''_{\rm nl}(\bar{\phi})$ leads to growth of perturbations of the form $\delta\phi_\bk\sim P_{\bk}(t)e^{\Re[\tilde{\mu}_k]t}$ where $\tilde{\mu}_k$ are the Floquet exponents and $P_{\bk}(t)$ are periodic functions. We find that $\mu_k\approx\Re[\tilde{\mu}_k]$ for $\bar{\phi},\bar{\psi}\ll 1$ and $k\lesssim 1$. The boundary of the non-relativistic band yields a good approximation to the relativistic case for $\bar{\psi}\lesssim1$.

\subsection{Non-relativistic Solitons and Oscillons}

It is worth making a comparison of our non-relativistic solitons discussed in Section~\ref{sec:SSI} to the relativistic ones (oscillons). Recall that $\phi=\sqrt{2}\Re[\psi e^{-i t}]$. For small amplitude solitons (oscillons), we expect $\phi(t,r)\approx \phi(r)\cos(\omega t)+\hdots$ (with $\omega<1$). The solitons in the nonlinear Schr\"{o}dinger equation have the form $\psi(t,r)=\Psi(r) e^{-i\nu t}$. Hence
\Beq
\phi(r)\cos[\omega t]\approx \sqrt{2}\Psi(r)\cos[(1+\nu)t]\qquad \Longrightarrow\qquad\Psi(r)\approx\frac{1}{\sqrt{2}}\phi(r)\,\qquad\nu \approx \omega-1\,,
\Eeq
where $\nu<0$. We caution the reader that this small amplitude analysis should merely be taken as a guide.  The actual relativistic solitons can include multiple frequencies, including breathing modes at large amplitudes. We compared the profiles of relativistic solitons obtained from the simulations in \cite{Lozanov:2019ylm}, and found good qualitative agreement with Fig. \ref{fig:sol}, albeit with more scatter around the curve (after appropriate scaling of the parameters). 

A numerical study of the lifetime and stability of large amplitude relativistic oscillons (but without gravitational interactions) in flattened potentials like the one we use here has been discussed in \cite{Salmi:2012ta}. A more detailed connection between non-relativistic solitons and oscillons, as well as analysis of the stability of relativistic cases (typically for small amplitude) can be seen in \cite{Fodor:2008es,Hertzberg:2010yz,Mukaida:2016hwd,Saffin:2014yka}. 

\subsection{Probability Density Functions of the Density and Gravitational Potential}
\begin{figure*}[h!]
\begin{tabular}{cc}
\includegraphics[width=0.43\textwidth]{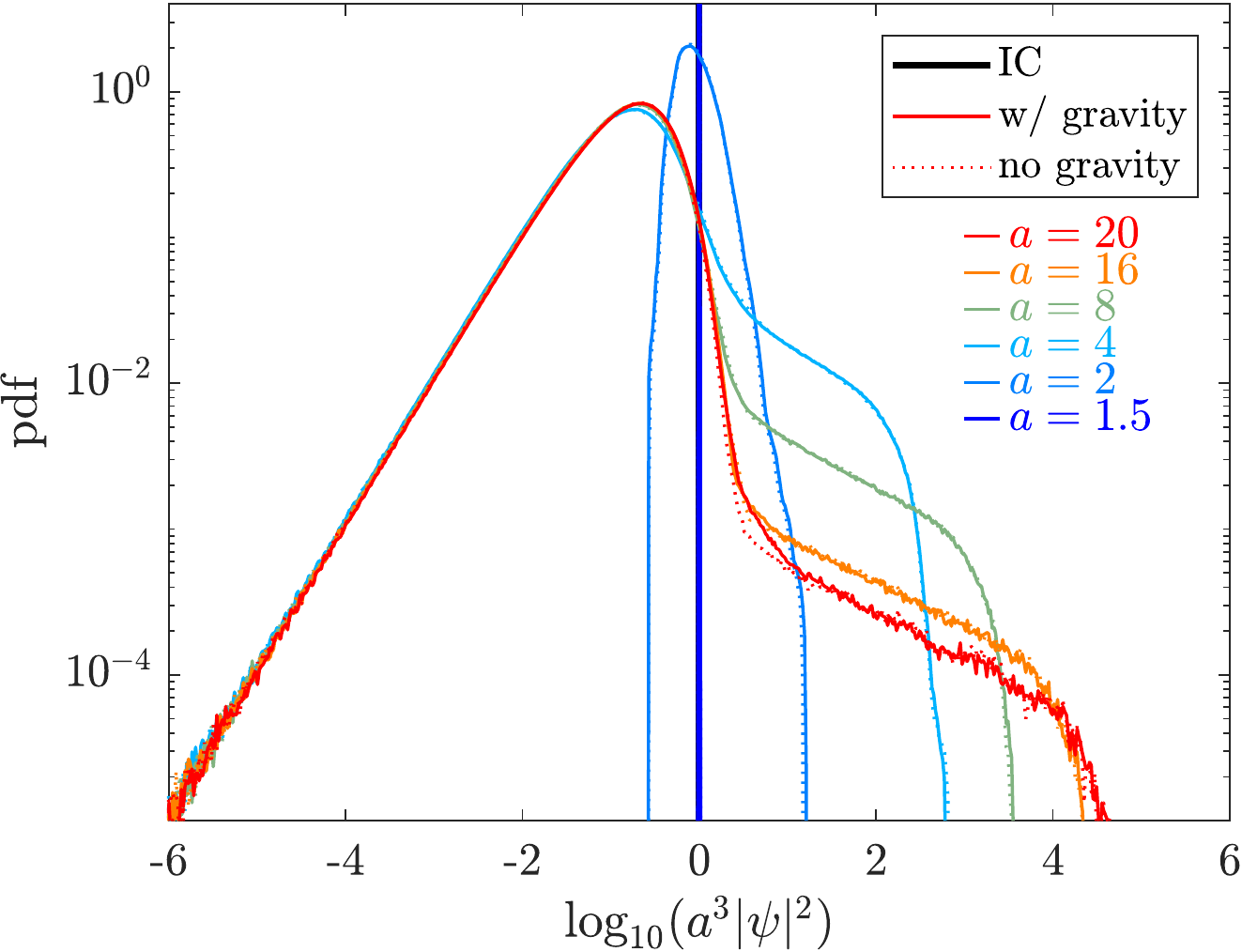}  & \qquad
\includegraphics[width=0.445\textwidth]{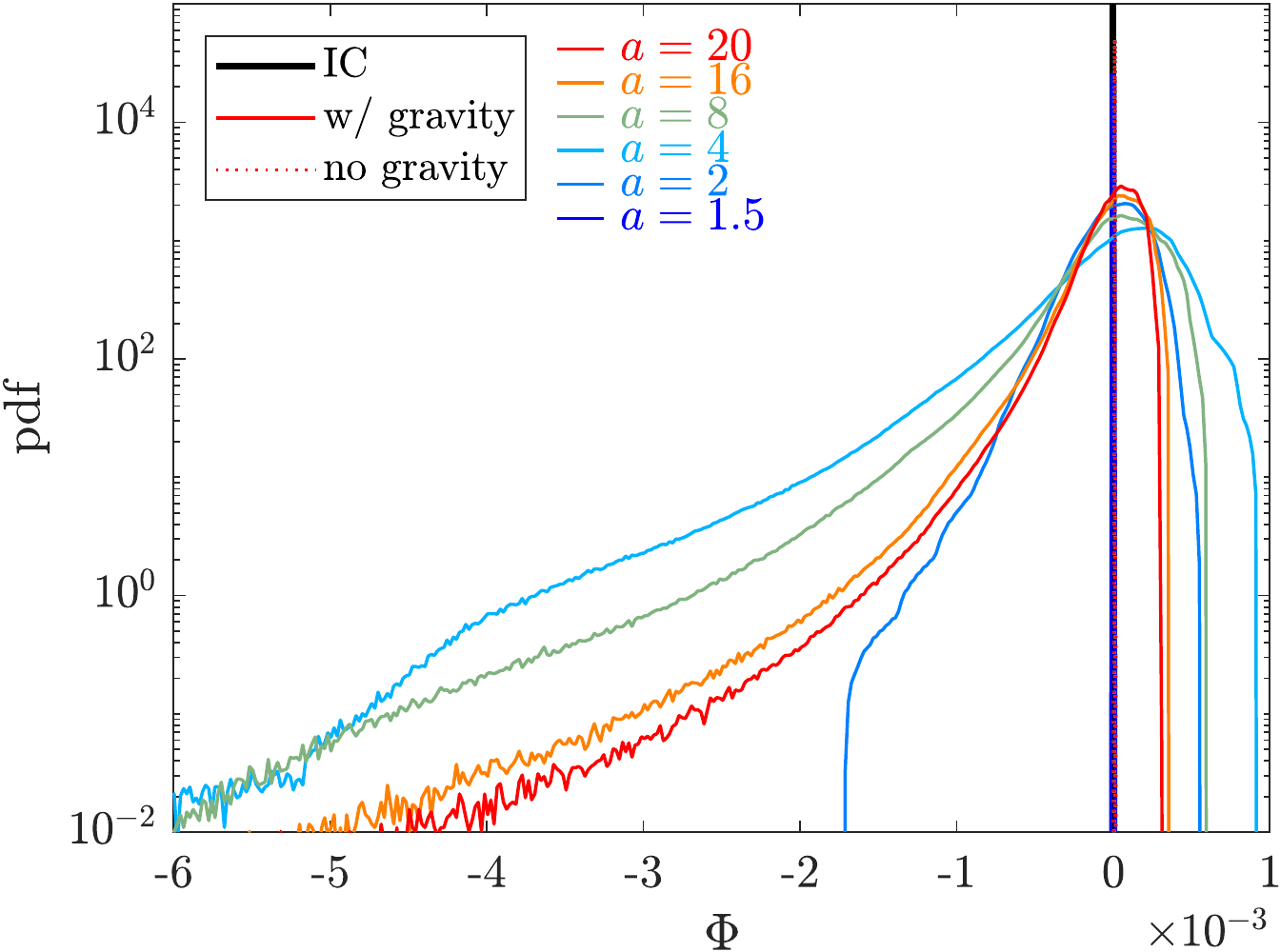}  
\end{tabular}
\label{fig:nzeta}
\caption{Probability density function of density of the field (left panel)  and the gravitational potential (right panel). The PDF of the density is shown for the case with and without gravitational interactions included. In the PDF for the gravitational potential, at each time slice the spatial average of the gravitational potential is zero in the simulation volume. Note that the gravitational potential remains small throughout our simulation. The behavior of the density PDF here can be compared to simulations which involve the relativistic Klein-Gordon equation in an expanding universe but with a ``passively'' calculated gravitational potential (Fig. 3 of \cite{Lozanov:2019ylm}).}
\end{figure*}

The probability density function (PDF) of the energy density and the gravitational potential in our simulation is show in Fig.~\ref{fig:nzeta}. Note that the gravitational potential in the simulation volume remains small $|\Phi|\ll 1$. Moreover the formation of the ``shelf'' in the density PDF($a\gtrsim 4$) is characteristic of the systems in which soliton formation takes place; the same qualitative behavior was seen when simulating relativistic systems with a related self-interaction potential \cite{Lozanov:2019ylm} (see Fig 3. in that paper; however, note that $\beta\approx 8\times 10^{-3}$ in that figure). Note that $\beta\ll 1$ is required for the instability that generates solitons to be effective in a self-consistently expanding universe. The same $\beta$ also controls the strengths of the gravitational potential. This competition makes it difficult to generate individual solitons with large gravitational potentials via the self-interaction instability.

\end{document}